\documentclass[10pt]{article}
\usepackage[utf8]{inputenc}
\usepackage{geometry}
\usepackage{amsmath,amssymb,amsthm,mathrsfs,nicefrac}
\usepackage{mathtools}
\usepackage{latexsym,amscd,amsbsy,dsfont,amsfonts}
\usepackage{graphicx,wrapfig}
\usepackage{authblk,textcomp}
\usepackage[dvipsnames]{xcolor}
\usepackage{cancel}

\usepackage{bm}
\usepackage[cal=euler]{mathalfa}
\usepackage{libertine}
\usepackage[libertine,smallerops]{newtxmath}
\usepackage[T1]{fontenc}
\usepackage[font=small]{caption}
\usepackage{tikz}
\usetikzlibrary{arrows}
\usetikzlibrary{decorations}
\usepackage{subcaption}
\usepackage{booktabs}
\usepackage{doi}
\usepackage{float}
\usepackage{enumitem}
\usepackage{hyperref}
\hypersetup{pdfauthor={rveiga},pdftitle={rbm_flow},%
            colorlinks, linktocpage=true, pdfstartpage=1, pdfstartview=FitV,%
            breaklinks=true, pdfpagemode=UseNone, pageanchor=true, pdfpagemode=UseOutlines,%
            plainpages=false, bookmarksnumbered, bookmarksopen=true, bookmarksopenlevel=1,%
            hypertexnames=true, pdfhighlight=/O,%
            urlcolor=orange, linkcolor=blue, citecolor=NavyBlue}

\usepackage[toc,page]{appendix} 
\appto\appendix{\counterwithin{equation}{section}}

\title{Restricted Boltzmann Machine Flows and The Critical Temperature of Ising models}
\author[1]{Rodrigo Veiga}
\author[2,3]{Renato Vicente}
\affil[1]{\small Universidade de S\~{a}o Paulo. Instituto de F\'{i}sica. S\~{a}o Paulo, SP, Brazil.}
\affil[2]{\small LatAm Experian DataLab. S\~{a}o Paulo, SP, Brazil.}
\affil[3]{\small Universidade de S\~{a}o Paulo. Instituto de Matem\'{a}tica e Estat\'{i}stica. S\~{a}o Paulo, SP, Brazil.}
\date{}

\begin{document}

\maketitle

\begin{abstract}

We explore alternative experimental setups for the iterative sampling (flow) from Restricted Boltzmann Machines (RBM) mapped on the temperature space of square lattice Ising models by a neural network thermometer. This framework has been introduced  to explore connections between RBM-based deep neural networks and the Renormalization Group (RG). It has been found that, under certain conditions, the flow of an RBM trained with Ising spin configurations approaches in the temperature space a value around the critical one: $ k_B T_c / J \approx 2.269$. In this paper we  consider datasets with no  information about  model topology to argue that a neural network thermometer is not an accurate way to detect whether the RBM has learned scale invariance or not.
 \end{abstract}

\tableofcontents

\section{\label{sec:intro} Introduction}

The observation that Neural Networks can be studied by Statistical Physics techniques is not new \cite{engel_2001}. Following the Deep Learning revolution, the past few years have also witnessed a boost of activity on the applications of Machine Learning (ML) algorithms as a tool to study complex physical models. These include phase identification in both classical and quantum systems \cite{wang_2016, carrasquilla_2017, cheng_2017}, dimensionality reduction of a Hilbert space representing the wave function through reinforcement learning \cite{carleo_2017}, generative models applied to classical systems \cite{morningstar_2018, cossu_2019}, or even the development of new algorithms capable of finding coarsed-grained transformations \cite{koch-janusz_2018}; among many other examples. We now regard ML as both a useful numerical tool for doing physics \cite{zdeborova_2017, mehta_2019} as much as interesting physical systems themselves \cite{beny_2013}.

In particular, the relation between unsupervised learning based on standard Restricted Boltzmann Machines (RBMs) and the Renormalization Group (RG) in Kadanoff's picture \cite{kadanoff_2000}, pointed out in a seminal paper by Mehta and Schwab \cite{mehta_2014}, has attracted some attention \cite{tegmark_2017,iso_2018}. 

For the purpose of bringing some light to the discussion, Iso {\it et al.} \cite{iso_2018} trained RBMs using Monte Carlo (MC) samples from square lattice ferromagnetic Ising models with homogeneous nearest neighbor exchange interactions $J>0$ and no external field. With the introduction of a standard classification neural network (NN) as a thermometer, they were able to map the probability flow from the trained RBM (samples from the trained model) with a flow in the Ising model parameters space. By monitoring flows of RBMs trained with a joint dataset of states with temperatures $T$ below and above the critical value $k_B T_c / J \approx 2.269$ \cite{onsager_1944}, these authors have observed that samples generated by the machine flow towards a stable fixed point around $T_c$. Although this behavior is opposite to the conventional RG flow \cite{goldenfeld_1992}, there is this interesting coincidence in the location of the fixed point.

In this paper, we seek to contribute to the understanding of why those scale invariant configurations would be attractors of the RBM flow. For that we begin by reproducing the main result of Ref. \cite{iso_2018}; namely, that the RBM flow goes towards a fixed point around $T_c$.
We then analyze an RBM trained with a multi-temperature set of states from the mean field (MF) Ising model \cite{pathria_2011}.  Since the MF dataset does not contain the correct information about spin nearest neighbor correlations, we would expect the flow not to converge to the same fixed point. However, it does. 

Next we consider RBM training with a dataset composed only of states with $T=0 $ and $ T = \infty$. This is also a paradigmatic case. As two-dimensional Ising states are fed to the machine as vectors, not matrices, the RBM has no information about lattice dimensionality. Still, using the same NN thermometer, we found that the flow still goes towards a value around $T_c \approx 2.269$ (henceforth we consider temperatures measured in units of $k_B / J$).

This set of experiments bespeak in favor of a misinterpretation of the temperature measurement. We argue that in some cases the information about the geometry of the spin system is actually on the measurement device and the flow towards the critical temperature may be artifactual.

Section \ref{sec:RBM_NN} briefly reviews  RBMs and introduces the ideas of the RBM flow and of the NN thermometer according to Ref. \cite{iso_2018}. Section \ref{sec:featureextraction} reproduces the main results in \cite{iso_2018}, extends them to larger systems and discuss the calibration of the NN thermometer. Section \ref{sec:alternative} presents a series of experimental setups where information about correct correlations of the model is not presented to the RBM. In Section \ref{sec:annealed} we show that the flow towards $T_c$ does not depend on the specific values of RBM couplings, but only on their distribution. Section \ref{sec:w_matrix_analysis} studies singular values and eigenvalues decomposition of weight matrices. Closing remarks are then presented in Section \ref{sec:conclusion}.

\paragraph{Reproducibility} Code on \href{https://github.com/rodsveiga/rbm_flows_ising}{GitHub}.

\section{ \label{sec:RBM_NN} RBMs and the Neural Network Thermometer}

\subsection{{\label{RBM}} The Restricted Boltzmann Machine}  

\begin{figure}[ht!]
\begin{center}
\centerline{\includegraphics[width=0.35\textwidth]{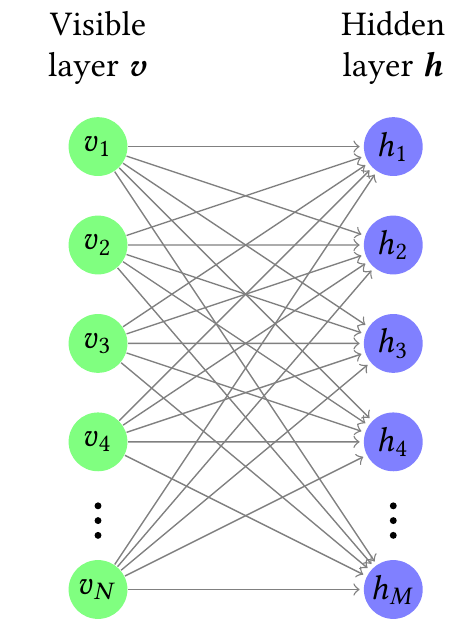}}
\caption{ An RBM with $N$ visible, $ \bm{v} = (v_1, v_2, ..., v_N) \in \mathbb{R}^N $, and $M$ hidden units, $ \bm{h} = (h_1, h_2, ..., h_M) \in \mathbb{R}^M$.}
\label{RBM_figure}
\end{center}
\vskip -0.2in
\end{figure}

An RBM \cite{smolensky_1986} is a generative model defined by a joint Boltzmann-Gibbs distribution $p( \bm{v}, \bm{h})$ with the following energy function:
\begin{equation}
\label{RBM_energy}
E\left(\bm{v}, \bm{h} \right)= - \sum_{j=1}^{N} \sum_{k=1}^{M} w_{jk} v_j h_k 
- \sum_{j=1}^{N} a_{j} v_{j} 
- \sum_{k=1}^{M} b_k h_k   \;,
\end{equation}
where $v_j$ denotes the state of the $j$-th visible unit and $h_k$ the state of the $k$-th hidden unit. The weight matrix $\bm{W} \in \mathbb{R}^{N\times M}$ is composed of elements $w_{jk}$ connecting neurons with labels $j$ and $k$. External fields acting on visible and hidden units are denoted, respectively, by $a_j$ and $b_k$. The graphical representation of an RBM is depicted in Figure \ref{RBM_figure}.

As an RBM is represented by a bipartite graph, the hidden variables are independent given the visible variables and vice versa. Additionally, the conditional distributions factorize and block Gibbs sampling  \cite{resnik_2010} can be employed:
\begin{subequations}
\label{eq:trans_prob}
\begin{equation}
\label{v_to_h}
     p\left( \bm{h} | \bm{v} \right) = \prod_{k=1}^{M} p\left( h_k | \bm{v} \right) 
\;,
\end{equation}
\begin{equation}
  p\left( \bm{v} | \bm{h} \right) = \prod_{j=1}^{N} p\left( v_j | \bm{h} \right)
\;.
\label{h_to_v}
\end{equation}
\label{eq:p_trans}
\end{subequations}

The training set $S = \{ \bm{v}^{(1)}, ...,  \bm{v}^{(l)} \}$ is independently generated by some (generally unknown) probability distribution $r( \bm{v} )$ and the learning process chooses the parameters $ \bm{\Theta} \equiv \{ w_{jk}, a_j, b_k  \} $ in order to minimize the KL-divergence \cite{cover_2006} between $r( \bm{v} )$ and $p( \bm{v} )$:
\begin{equation}
\label{KL}
D_{KL}\left(r\left(\bm{v}\right) \Vert p\left( \bm{v} \right)\right)=
\sum_{\{\bm{v}\}}r\left(\bm{v}\right)\log\frac{r\left(\bm{v}\right)}{p\left(\bm{v}\right)} \;,
\end{equation}
where  $p\left(\bm{v}\right) = \sum_{ \{ \bm{h} \} } p\left(\bm{v}, \bm{h}\right) $. This is equivalent to maximizing 
the log-likelihood \cite{fischer_2014}
\begin{equation}
\label{log_likelihood}
\log {\cal L} \left( \bm{\Theta} \vert S \right)  =  \log  \prod_{j=1}^{l} p ( \bm{v}^{(j)} | \bm{\Theta} ) =  \sum_{j=1}^{l} \log p ( \bm{v}^{(j)} | \bm{\Theta} )  \;.
\end{equation}
Calculating the derivatives of Eq.(\ref{log_likelihood}) with respect to the parameters  we find:
\begin{subequations}
\label{eq:whole_grad}
\begin{eqnarray}
\sum_{ \bm{v} \in S }
\frac{\partial \log  {\cal L} \left( \bm{\Theta} |  \bm{v} \right) }{\partial{w_{jk}}} 
\propto \left\langle v_j h_k \right\rangle_{\text{data}} - \left\langle v_j h_k \right\rangle_{\text{model}}
\;,
\label{lik_grad01}
\end{eqnarray}
\begin{equation}
\sum_{ \bm{v} \in S }
\frac{\partial \log {\cal L} \left( \bm{\Theta} |  \bm{v} \right) }{\partial a_j } 
\propto \left\langle v_j \right\rangle_{\text{data}} - \left\langle v_j \right\rangle_{\text{model}}
\;,
\label{lik_grad02}
\end{equation}
\begin{equation}
\sum_{ \bm{v} \in S }
\frac{\partial \log {\cal L} \left( \bm{\Theta} |  \bm{v} \right) }{\partial b_k } 
\propto \left\langle h_k \right\rangle_{\text{data}} - \left\langle h_k \right\rangle_{\text{model}}
\;,
\label{lik_grad03}
\end{equation}
\end{subequations}
where $\left\langle ... \right\rangle_{\text{data}}$ represents the expectation over the distribution $p(\bm{h}| \bm{v}) q(\bm{v})$, with $q(\bm{v})$ being the empirical distribution. Analogously, $\left\langle ... \right\rangle_{\text{model}}$ stands for the expectation over the model distribution $ p ( \bm{v}, \bm{h})$. 

Summing over all visible or hidden variables is intractable. Methods to tackle the expectation over the model, such as contrastive divergence (CD) learning \cite{hinton_2002}, parallel tempering \cite{desjardins_2010} and persistent contrastive divergence \cite{tieleman_2008} are available. In this paper, we have used CD, which has become a standard way to train RBMs. Instead of approximating the second term in the log-likelihood gradient using
samples from the model distribution, CD uses a Gibbs chain run for only $k $ steps (usually $k=1$ is enough) and initialized with an element $\bm{v}_{0} $ of the training set $S$, yielding the sample $\bm{v}_{k}$ after $k$ steps. Each step $t$ consists of sampling $\bm{h}_t$ from Eq.(\ref{v_to_h}) and subsequently sampling $\bm{v}_{t+1}$ from Eq.(\ref{h_to_v}). After $k$ steps, the expectations over the model distribution in Eqs.(\ref{eq:whole_grad}) are approximated by an expectation over $ p ( \bm{h} | \bm{v}_k) $.

\subsection{{\label{sec:RBM_flow}} The RBM flow}  

\begin{figure}[ht!]
\begin{center}
\centerline{\includegraphics[width=0.7\textwidth]{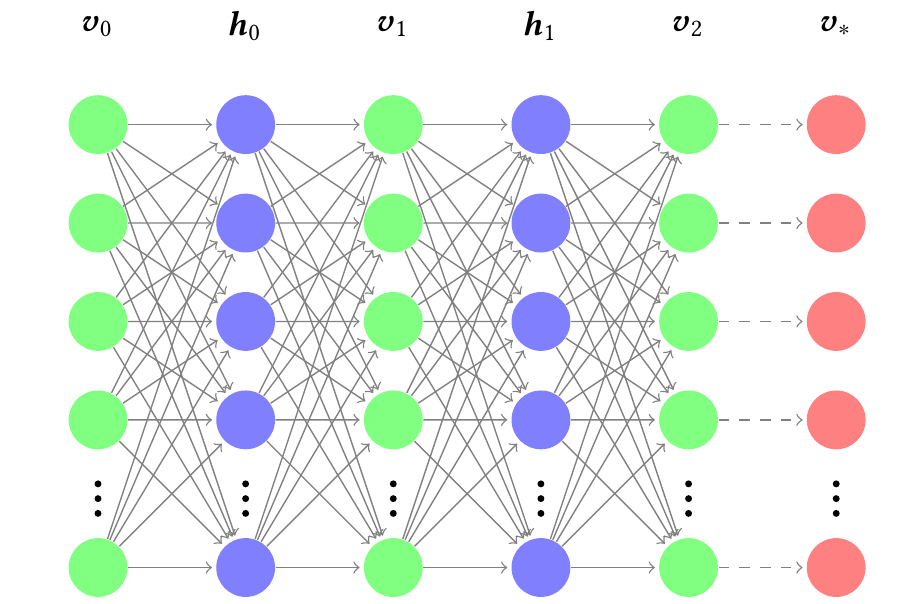}}
\caption{RBM flow for a machine with $N$ visible and $M$ hidden units, which can also be viewed as multilayer neural network with fixed coupling parameters.}
\label{FLOW_figure}
\end{center}
\vskip -0.2in
\end{figure}

Once the model is trained, the RBM flow is obtained by sequentially sampling hidden variables given visible variables and vice-versa, producing the following Markov chain 
\begin{equation}
\label{RBM_flow}
\bm{v}_0 \rightarrow \bm{h}_0 \rightarrow \bm{v}_1 \rightarrow \bm{h}_1 \rightarrow
\bm{v}_2 \rightarrow \bm{h}_2    \rightarrow ...
\rightarrow \bm{v}_{*} \sim r   \;.     
\end{equation}
which approximates $r (\bm{v} )$ at equilibrium. This Markov chain can be represented by a graph as depicted in Figure \ref{FLOW_figure}. In  \cite{iso_2018, funai_2020}  the authors measure the temperature at each iteration in the visible layer of the RBM flow using a NN as a thermometer.

\subsection{{\label{NN}} The NN thermometer} 

The thermometer consists of a feedforward NN classifier with a softmax output layer trained in a dataset composed by two dimensional $L \times L$ Ising microstates labelled by $K$ temperature values,
\begin{equation}
    \label{eq:NN_therm_scale}
{\cal T} = \{  T^{(1)}, T^{(2)}, ..., T^{(K)}   \} \;.
\end{equation}
Observe that the choice of ${\cal T}$ is arbitrary  and defines the scale of the thermometer, which is calibrated by cross-entropy minimization \cite{goodfellow_2016}. 

After training (calibration), the NN thermometer can therefore be used to attribute a probability for the temperature of a given sample configuration. An estimate for the temperature can be obtained by averaging over many samples at an unknown temperature. Assuming that the temperature of a set of microstates is provided by the most probable value of this empirical probability distribution, the NN thermometer can translate \eqref{RBM_flow} to a Markov chain dynamics in the temperature space,
\begin{equation}
    \label{eq:temp_flow}
    T_0 \rightarrow T_1 \rightarrow T_2 \rightarrow  ...
\rightarrow T_{*} 
\;.
\end{equation}
where all the measures are taken in the visible layer of the RBM. Hereafter we use $\nu$ to index the element $T_\nu$ of the flow dynamics given by \eqref{eq:temp_flow}.

\subsubsection{{\label{sub:sub:NN_calibration}} Calibration}

We calibrate two NN thermometers using MC spin configurations sampled from the Ising model with nearest neighbour interactions in a square lattice $L\times L$.

\begin{figure*}[htb!]
\centering
\begin{subfigure}[h]{.45\textwidth}
  \centering
  \includegraphics[width=\linewidth]{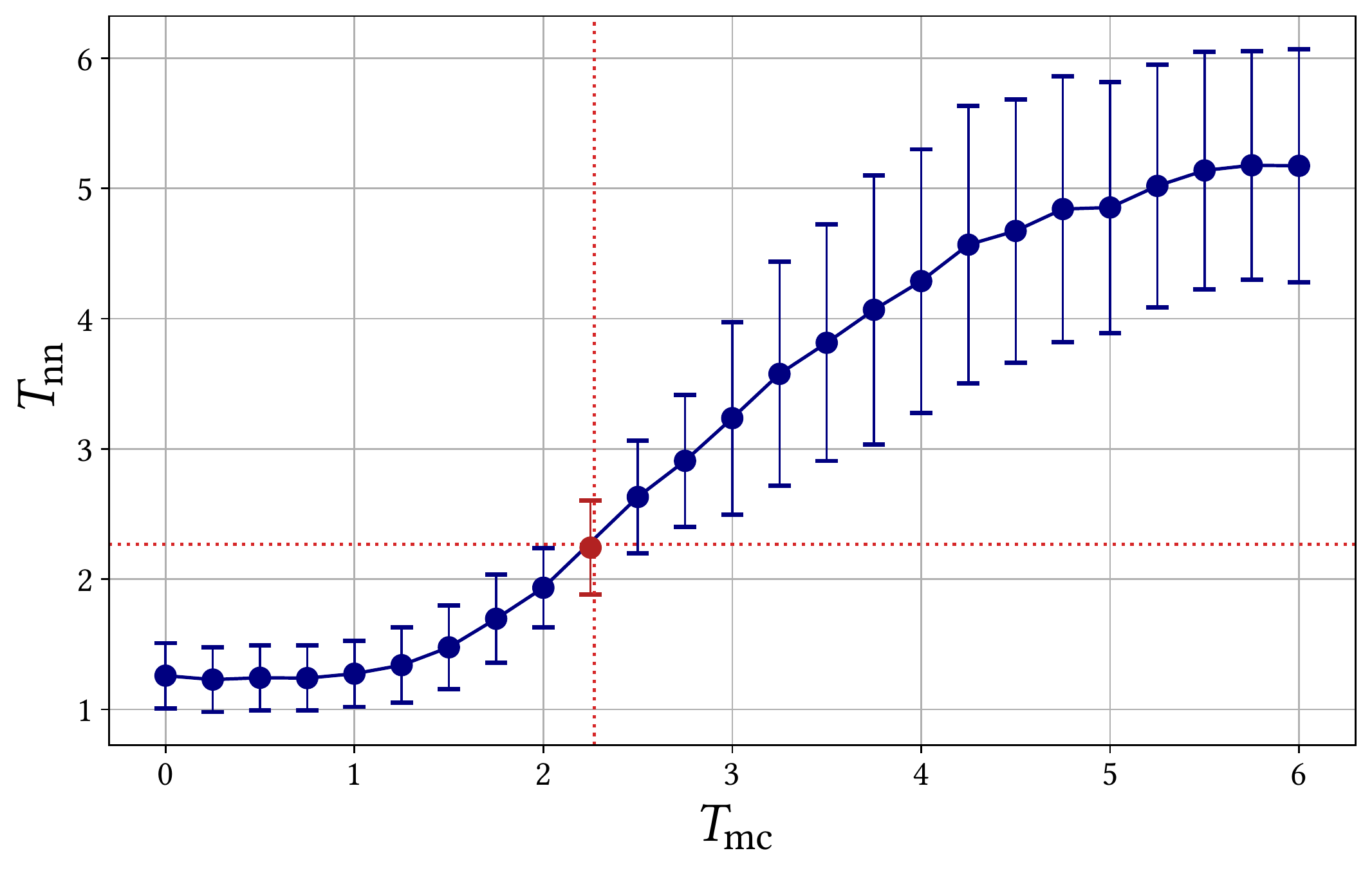}
  \caption{Calibration curve of the neural network thermometer for $L=10$.}
  \label{fig:L010_CALIB_THERM_Onsager}
\end{subfigure}%
\hspace{0.5cm}
\begin{subfigure}[h]{.45\textwidth}
  \centering
  \includegraphics[width=\linewidth]{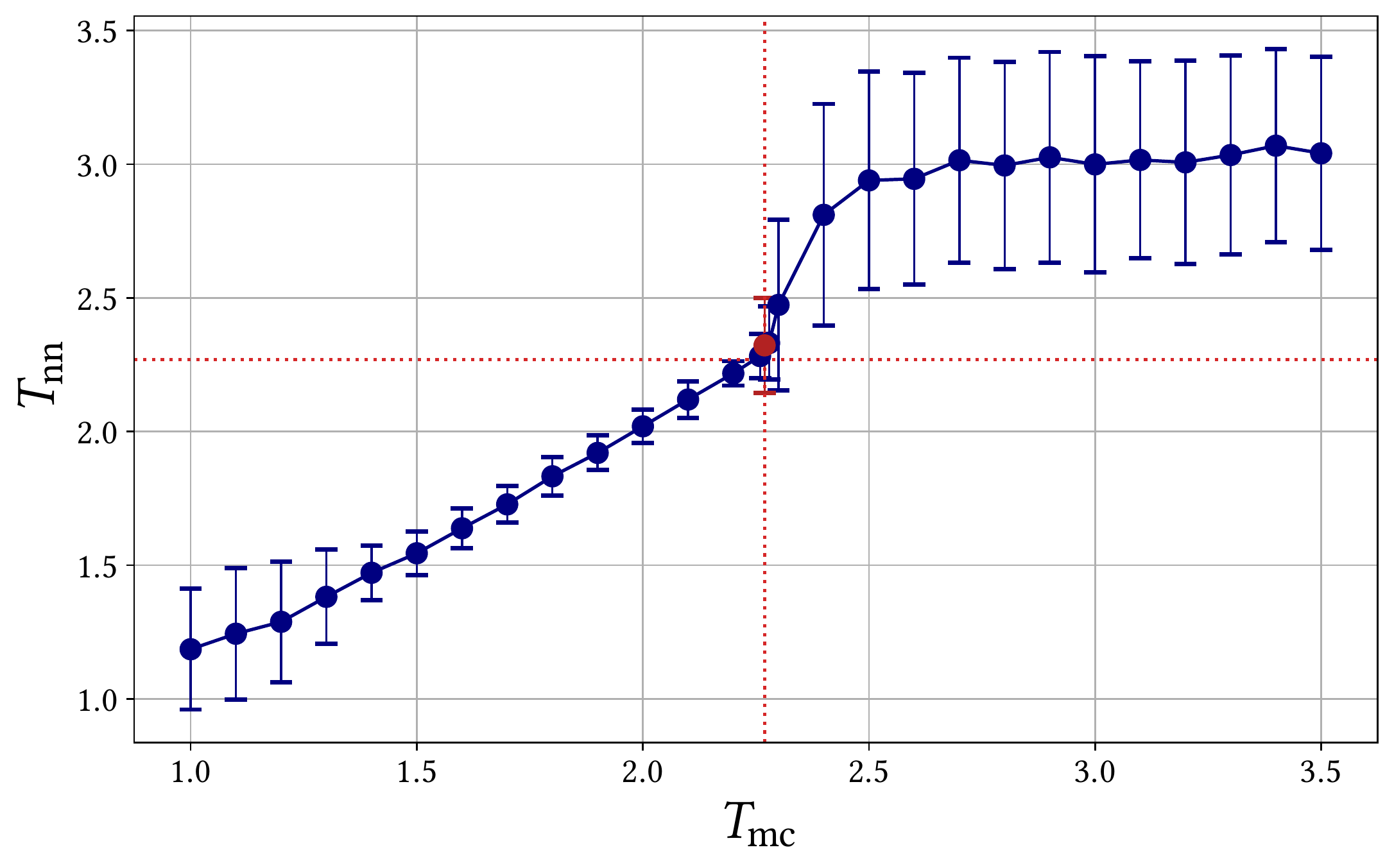}
  \caption{Calibration curve of the neural network thermometer for $L=100$}
 \label{fig:L100_CALIB_THERM_Onsager}
\end{subfigure}
\caption{Neural network temperature predictions $T_{\text{nn}}$ versus the true temperatures $T_{\text{mc}}$. Red point and red dashed lines: highlight the measures near $T_c$. Error bars: uncertainty within one standard deviation (each point is obtained from a average over the measures in a particular set of states).}
\label{fig:CALIB_THERM_Onsager}
\end{figure*}

For $L=10$, the dataset is composed by 2000 configurations for each of 25 different target temperatures: 
\begin{equation}
\label{eq:NN_scale_L010}
{\cal T}_{V}^{(L=10)} = \{ 10^{-6}, 0.25, 0.5, ..., 5.5, 5.75, 6 \} \;.
\end{equation}
The training set is constructed with 1800 states for each temperature and the test set with the remaining 200. After training the labelled test set is used to draw a calibration curve between the true temperature values $T_{\text{mc}}$ and the neural network predictions $T_{\text{nn}}$. This curve is presented in Figure \ref{fig:L010_CALIB_THERM_Onsager}. 

Notice that low temperatures are badly classified. The standard deviation also increases for temperatures higher than $T_c = 2.269$, in accordance with the results of \cite{iso_2018}, where sharp peaked empirical distributions were found for temperatures close to $T_c$. 

For $L=100$, the dataset is composed by 2000 configurations for each of 29 different target temperatures \footnote{The values around the critical temperature were included to increase accuracy in the calculation of thermodynamic variables, which are a sanity check of MC simulations. Configurations near $T_c$ are not, however, required to the RBM flows go towards $T_c$, as checked in Ref. \cite{iso_2018}}: 
\begin{equation}
\label{eq:NN_scale_L100}
{\cal T}_{V}^{(L=100)} = \{ 1, 1.1, ..., 2.259, 2.269, 2.279, ...,  3.4, 3.5\} \;.
\end{equation}
The calibration curve for $L=100$ curve is presented in Figure \ref{fig:L100_CALIB_THERM_Onsager}.

Throughout this work temperature measures are performed by the thermometers calibrated in this section.

\section{\label{sec:featureextraction} Scale Invariant Feature Extraction of Neural Network and the Renormalization Group Flow}

Here reproduce the main result obtained in Ref. \cite{iso_2018} for $L=10$ and extend their analysis to $L=100$. Henceforth we consider $N=M= L^2$ \footnote{In Ref. \cite{iso_2018} the authors trained RBMs with  $M= 16, 36, 64, 81, 100, 225$ and $400$ hidden units. In our experiments we have checked their result: RBM flow goes to a fixed point around $T_c$ for $M \le N $. As pointed out by them, this probably happens because RBM with $N<M$ captures too much irrelevant information.} and also fix biases to zero in all simulations.

\begin{figure*}[htb!]
\centering
\begin{subfigure}[h]{.45\textwidth}
  \centering
  \includegraphics[width=\linewidth]{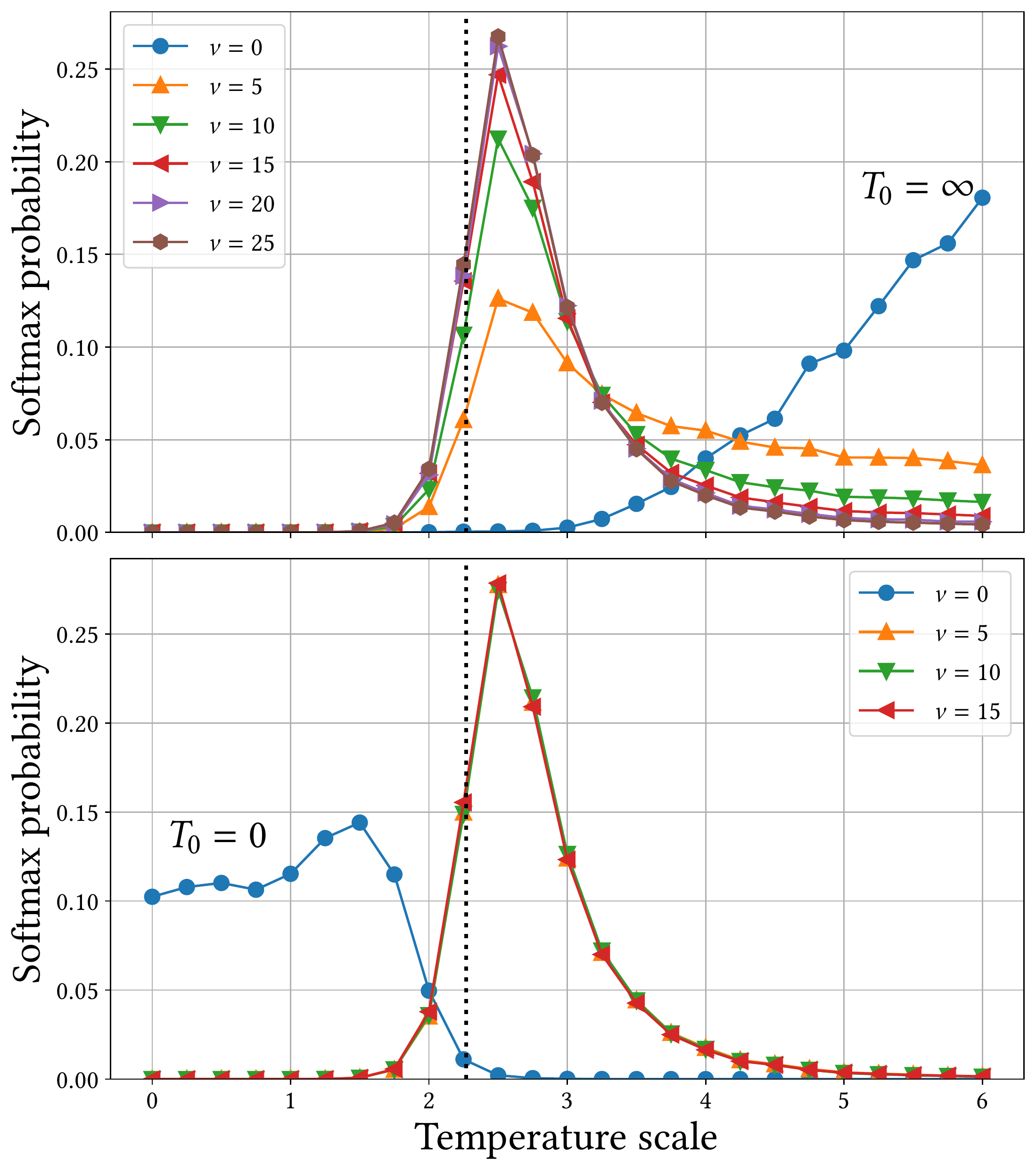}
  \caption{$L=10$: RBM trained with MC samples ${\cal T}_{V}^{(L=10)}$.}
  \label{fig:L010_RBM_flow_ALL}
\end{subfigure}%
\hspace{0.5cm}
\begin{subfigure}[h]{.45\textwidth}
  \centering
  \includegraphics[width=\linewidth]{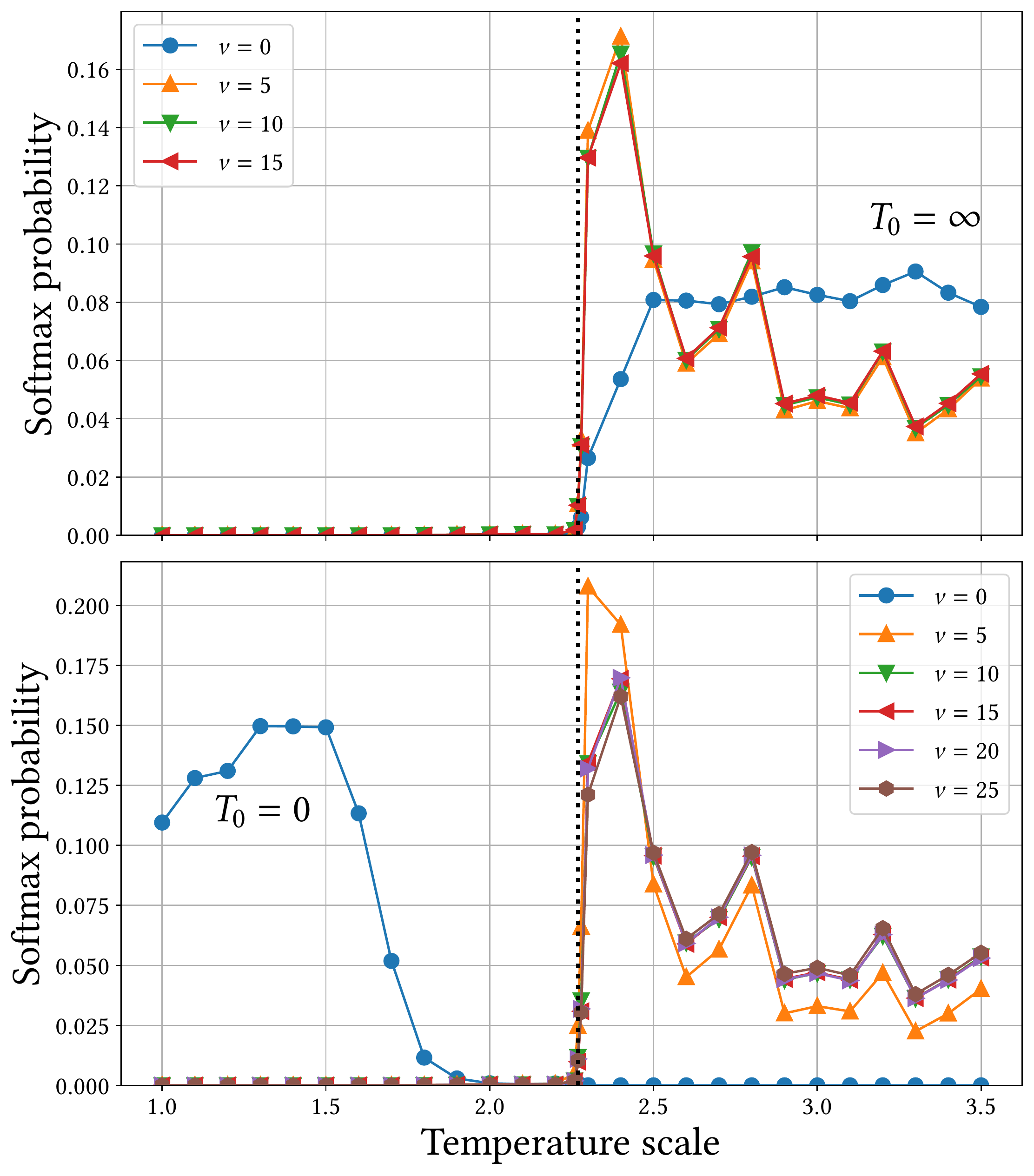}
  \caption{$L=100$: RBM trained with MC samples ${\cal T}_{V}^{(L=100)}$}
 \label{fig:L100_RBM_flow_ALL}
\end{subfigure}
\caption{RBM flows for multi-temperature datasets. Initial states: $T_0 = \infty$ on the top plots and $ T_0 = 0 $ on the bottom plots. Vertical dashed line: $T_c = 2.269$. Label $\nu$ refers to the element $T_\nu$ of the flow Markov chain \eqref{eq:temp_flow}.}
\label{fig:RBM_flow_ALL}
\end{figure*}

 In Figure \ref{fig:L010_RBM_flow_ALL} we present the flow for a machine trained with the dataset from Eq.(\ref{eq:NN_scale_L010}), the same one used to calibrate the thermometer. We verify flows towards a fixed point around $T_c$ whether the initial states $\bm{v}_0$ are sampled for $T_0 = \infty$ (random microstates)  or $T_0 = 0$ (ordered microstates). The same behavior is verified for $L=100$ in Figure \ref{fig:L100_RBM_flow_ALL}. Similarly, the machine was trained with the dataset from Eq.(\ref{eq:NN_scale_L100}).

\begin{figure*}[htb!]
\centering
\begin{subfigure}[h]{.45\textwidth}
  \centering
  \includegraphics[width=\linewidth]{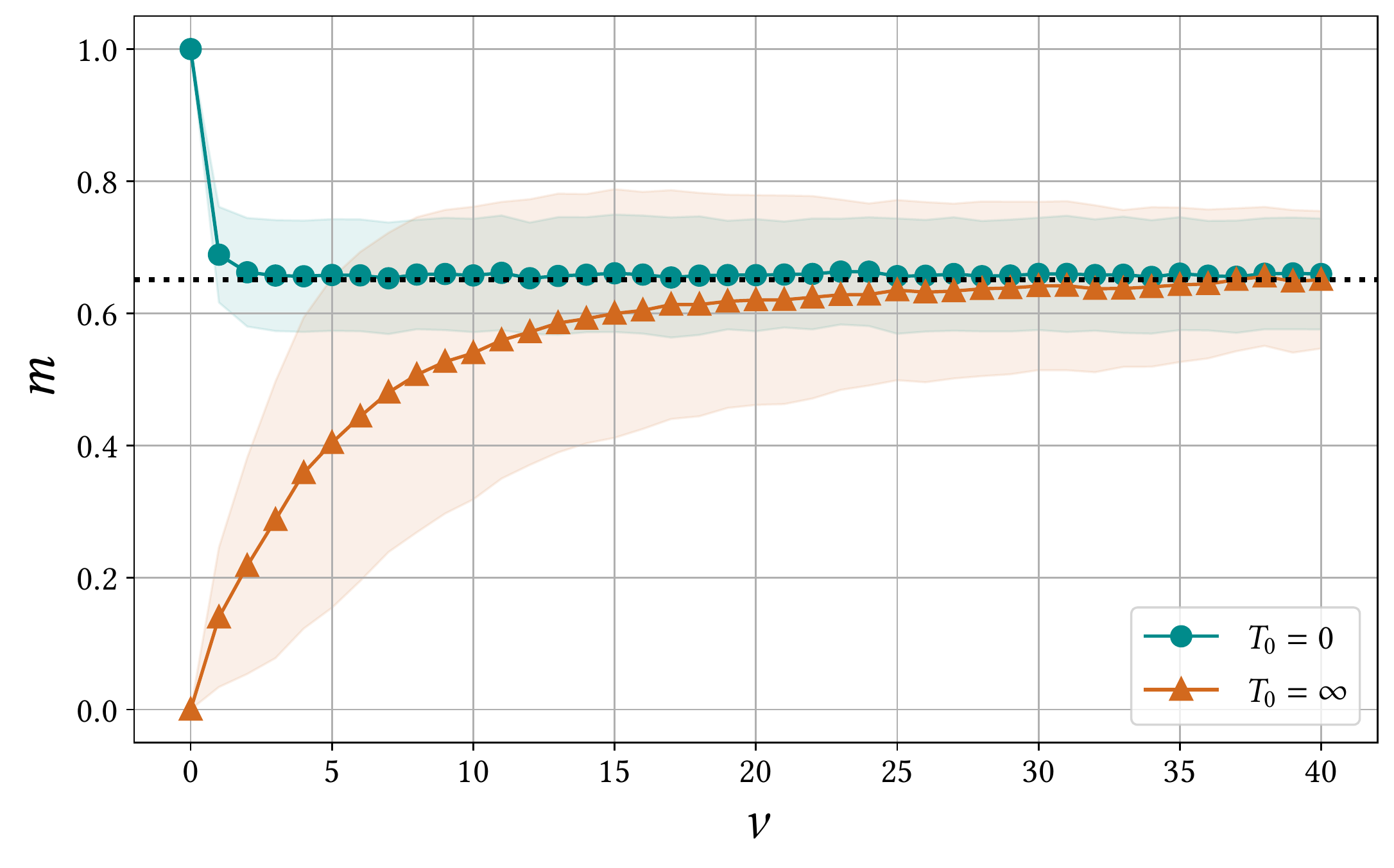}
  \caption{Magnetization $ m \equiv | \sum_{j}  v_j |  / N$ as a function of flow iteration ($L=10$) from two initial conditions: $T_0 = 0$ and $T_0 = \infty$. Horizontal dashed line: $m^{*} \approx 0.65$. Colored areas: uncertainty within one standard deviation. }
  \label{fig:L010_RBM_ALL_mag}
\end{subfigure}%
\hspace{0.5cm}
\begin{subfigure}[h]{.45\textwidth}
  \centering
  \includegraphics[width=\linewidth]{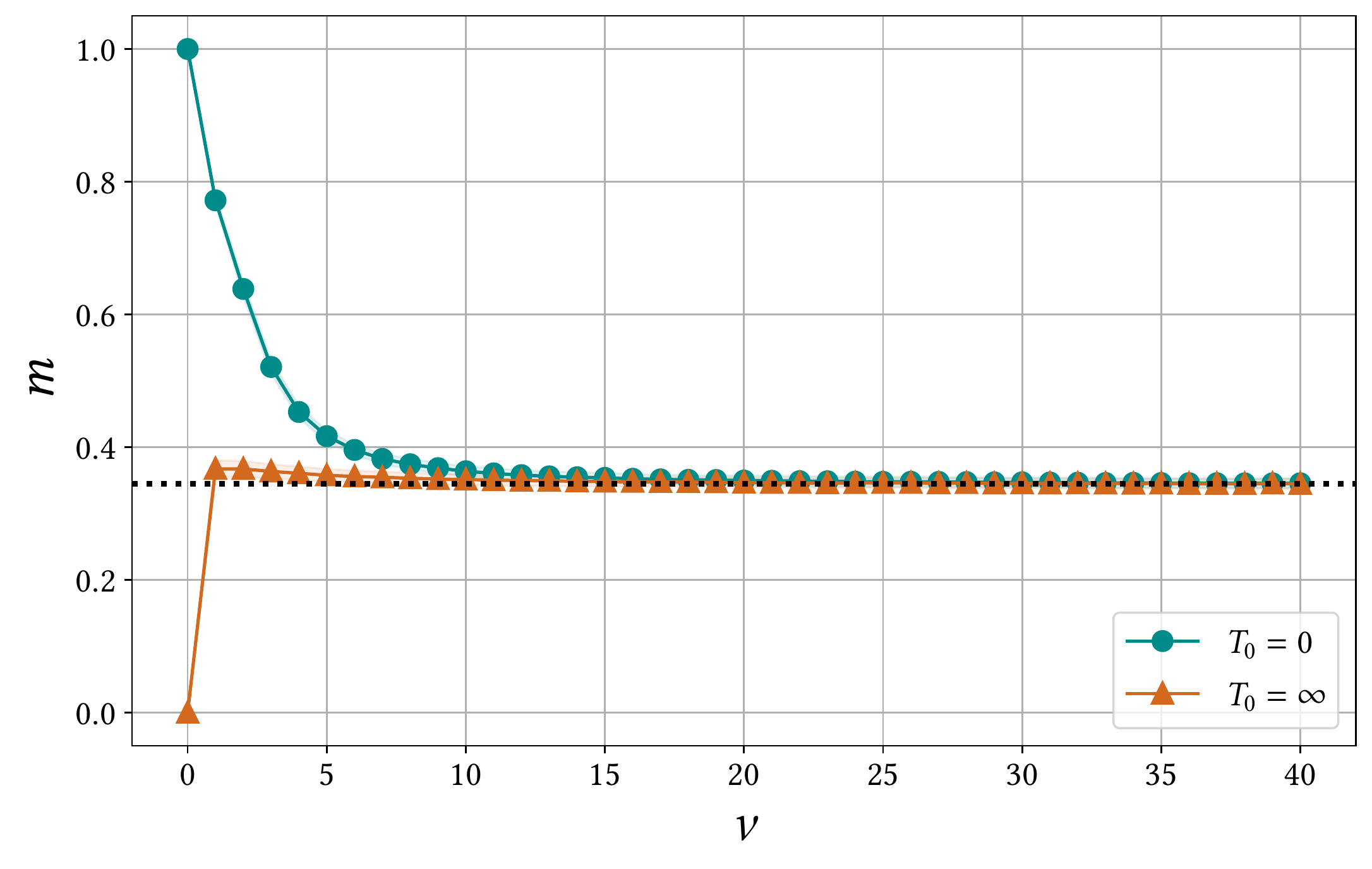}
  \caption{Magnetization $ m \equiv | \sum_{j}  v_j |  / N$ as a function of flow iteration ($L=100$) from two initial conditions: $T_0 = 0$ and $T_0 = \infty$. Horizontal dashed line: $m^{*} \approx 0.35$. Colored areas: uncertainty within one standard deviation.}
 \label{fig:L100_RBM_ALL_mag}
\end{subfigure}
\begin{subfigure}[h]{.45\textwidth}
  \centering
  \includegraphics[width=\linewidth]{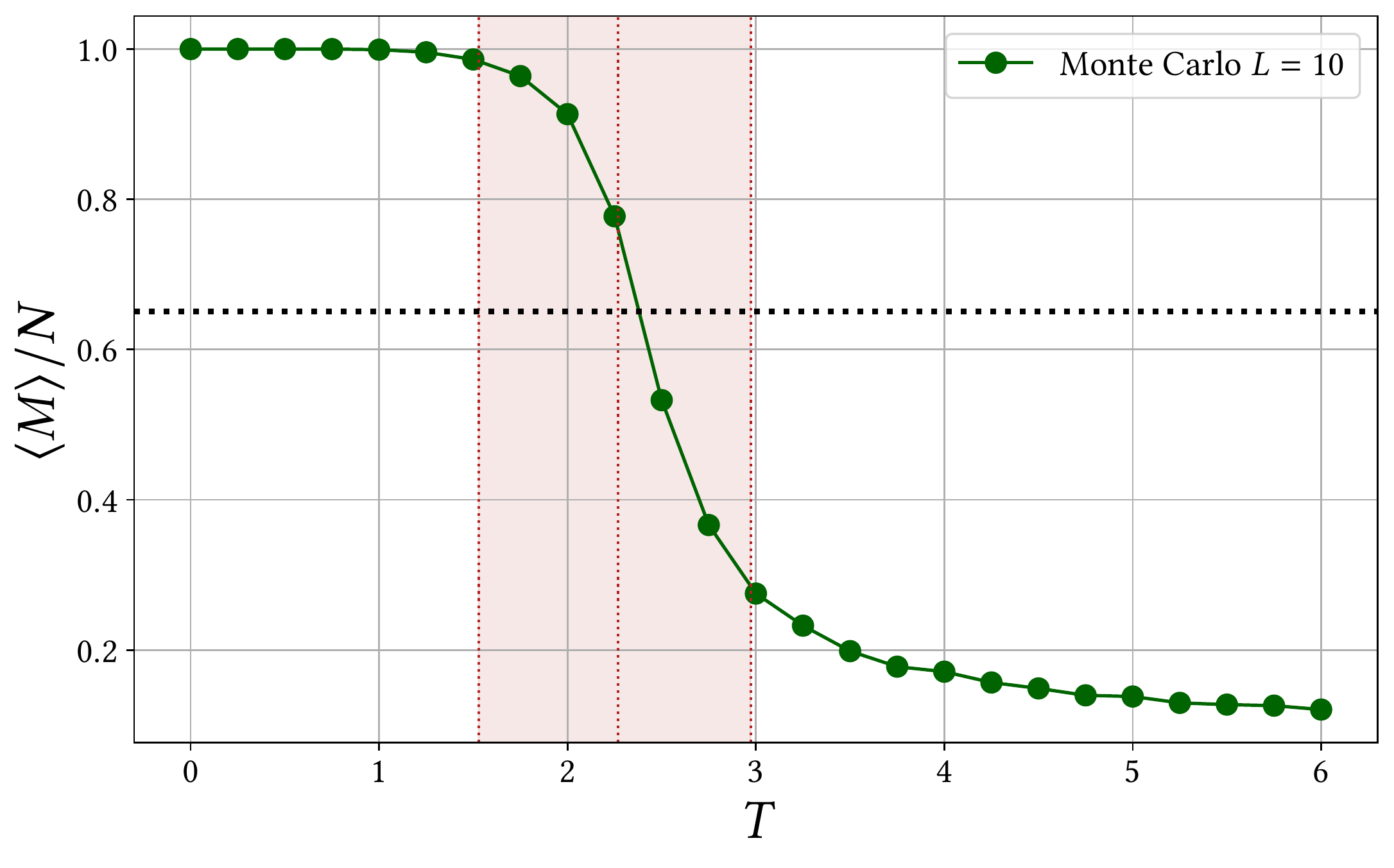}
  \caption{Magnetization per spin as a function of temperature obtained from the MC dataset ($L=10$) with which RBM and thermometer were trained. Horizontal dashed line: $m^{*} \approx 0.65$. Red filled area: projects the uncertainty of the thermometer around $T_c$ shown in Figure \ref{fig:L010_CALIB_THERM_Onsager} within two standard deviations (95\% confidence interval).}
 \label{fig:L010_RBM_flow_mag_nearTc_MC}
\end{subfigure}%
\hspace{0.5cm}
\begin{subfigure}[h]{.45\textwidth}
  \centering
  \includegraphics[width=\linewidth]{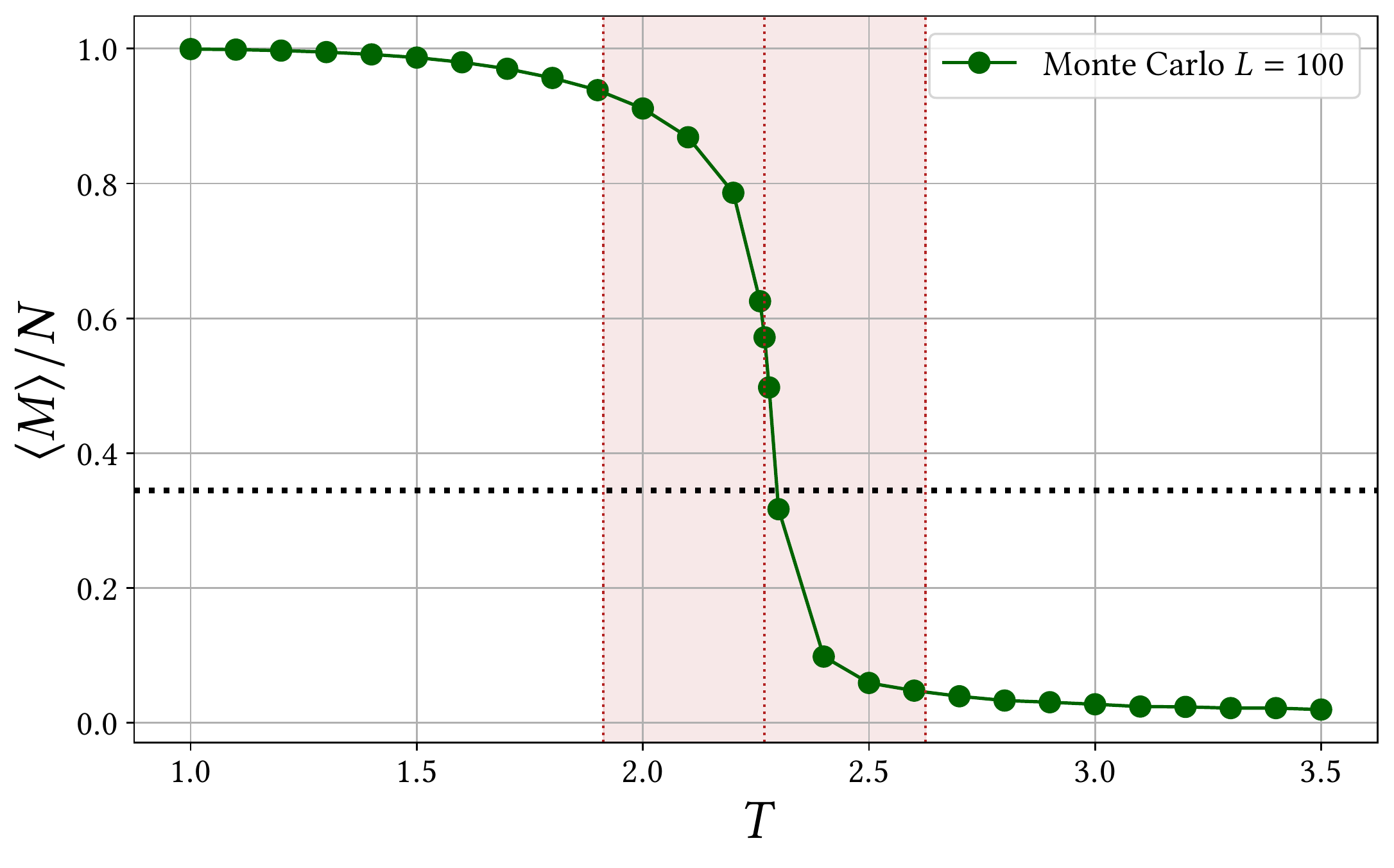}
  \caption{Magnetization per spin as a function of temperature obtained from the MC dataset ($L=100$) with which RBM and thermometer were trained. Horizontal dashed line: $m^{*} \approx 0.35$. Red filled area: projects the uncertainty of the thermometer around $T_c$ shown in Figure \ref{fig:L100_CALIB_THERM_Onsager} within two standard deviations (95\% confidence interval).}
 \label{fig:L100_RBM_flow_mag_nearTc_MC}
\end{subfigure}
\caption{RBM flows monitored from a magnetization-like function and their relation with the magnetization per spin as a function of temperature of the Monte Carlo datasets with which RBMs and thermometers were trained.}
\label{fig:RBM_flow_ALL_magflow_magMC}
\end{figure*}

We also inspect the  magnetization $ m \equiv | \sum_{j}  v_j |  / N$ through the RBM flow. For $L=10$, it can be seen in Figure \ref{fig:L010_RBM_ALL_mag} that the magnetization fixed point is $m^{*} \approx 0.65$. For $L=100$, $m^{*} \approx 0.35$ , as depicted in Figure \ref{fig:L100_RBM_ALL_mag}. 

As the fixed point does not depend on the flow initialization, it is clear that the dynamics either adds some level of disorder to initially ordered states or adds order to initially random states. Despite that, it is quite intriguing that states with such different order parameters are interpreted by the thermometer as states near critical temperature. In order to investigate that, we project the thermometer precision around $T_c$ into the magnetization $\langle M \rangle / N $ dependence of the Monte Carlo dataset. The results are presented in Figure \ref{fig:L010_RBM_flow_mag_nearTc_MC} for $L=10$ and in Figure \ref{fig:L100_RBM_flow_mag_nearTc_MC} for $L=100$. 

We call attention to the fact that a wide range of magnetization values would be equally compatible with a reading of $T_c$ by the NN thermometer. This observation encourages the investigation of alternative experimental setups in order to understand what is the relevant information to be fed to the RBM in order to produce the flow towards the critical temperature. What if the RBM were trained with samples with no information about  model topology?

\section{\label{sec:alternative} Alternative numerical experiments}

\subsection{ \label{sec:MF_training_set} Mean Field training set}

We now consider a dataset composed of spin states sampled from the model within the mean field (MF) approximation and no external field for $L=10$. Naturally, the correct spin-spin nearest neighbour correlations are not taken into account in this scenario, which does not predict the correct temperature at which the paramagnetic-ferromagnetic phase transition occurs. Within this approximation, we can easily solve the model in the thermodynamic limit to obtain from the Curie-Weiss equation: $T_{c}^{(MF)}= z$, where the coordination number $z$ is equal to the  lattice coordination number  (for example: $z=2$ for a 1D lattice; $z=3$ for a 2D triangular lattice; $z=4$ for a 2D square lattice or for a tetrahedral lattice; etc). 

\begin{figure*}[htb!]
\centering
\begin{subfigure}[h]{.45\textwidth}
  \centering
  \includegraphics[width=\linewidth]{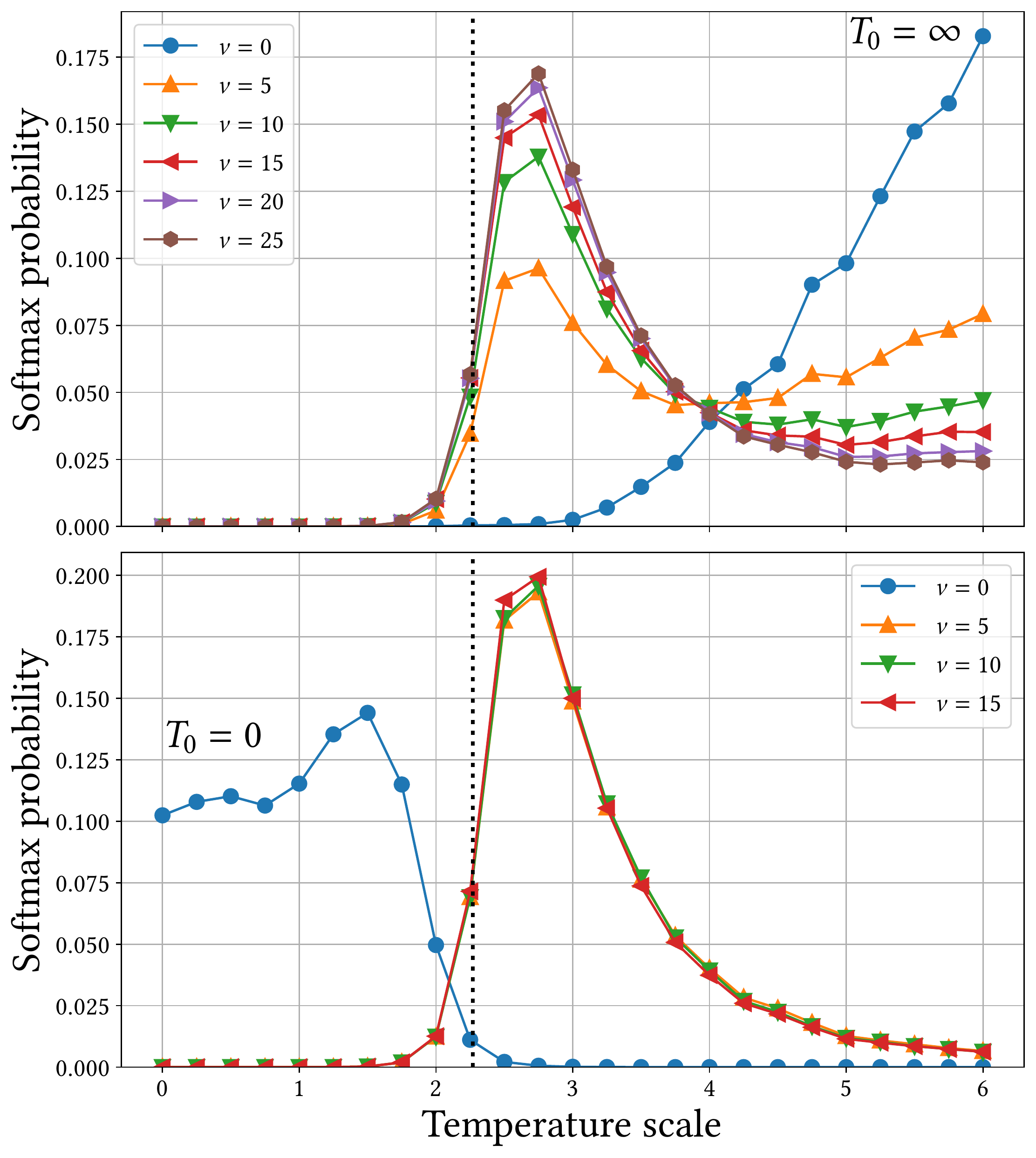}
  \caption{RBM flow for the MF dataset ($L=10$). Dataset set information: 2000 mean field spin configurations for each of the 25 temperatures from Eq.(\ref{eq:NN_scale_L010}).}
  \label{fig:L010_RBM_MF_FLOW}
\end{subfigure}%
\hspace{0.5cm}
\begin{subfigure}[h]{.45\textwidth}
  \centering
  \includegraphics[width=\linewidth]{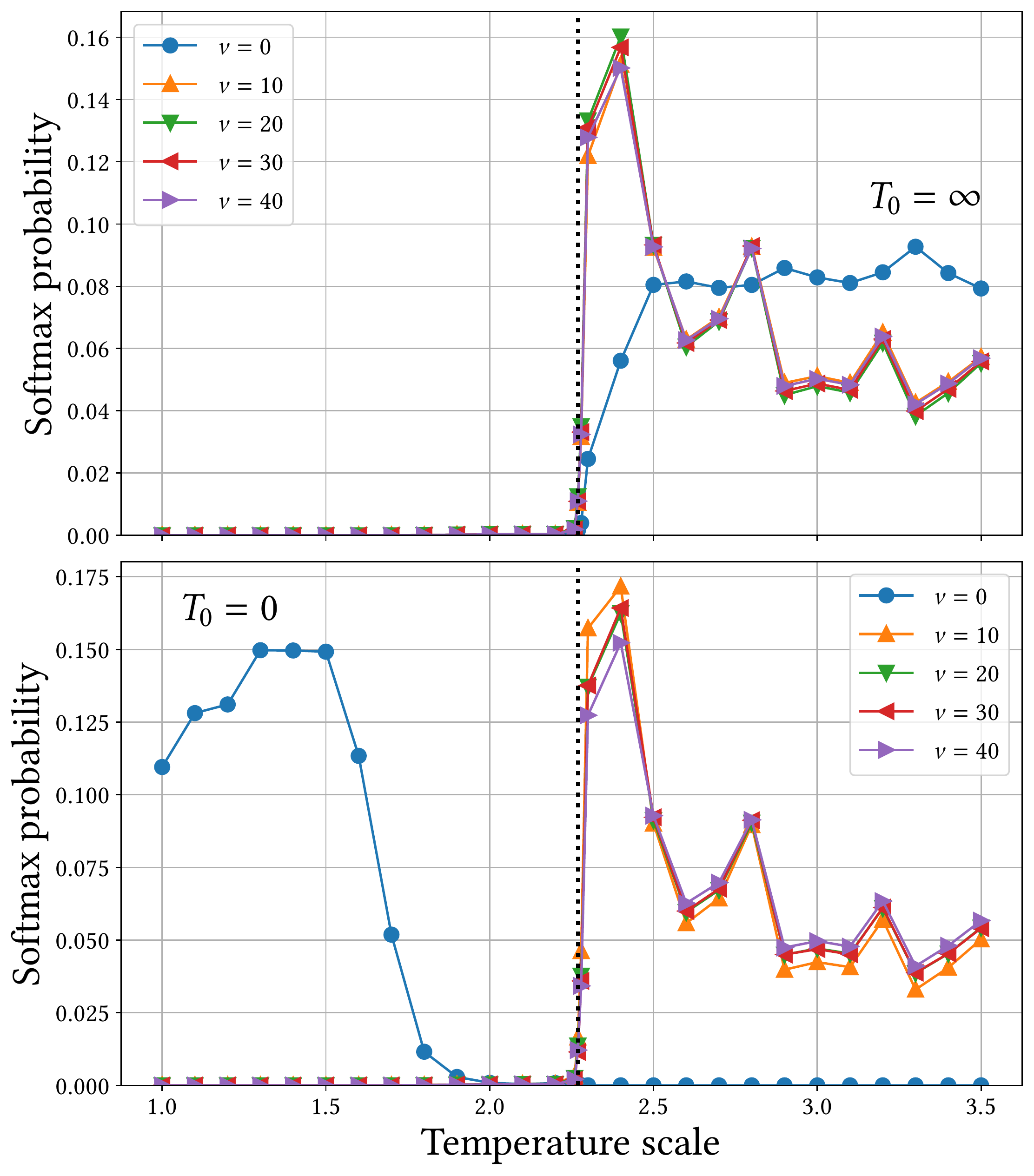}
  \caption{RBM flow for the $T=0$/$T=\infty$ dataset ($L=100$).}
 \label{fig:L100_RBM_ONLY_LOWHIGH}
\end{subfigure}
\caption{Alternative numerical experiments for the RBM flow proposed in in Section \ref{sec:alternative}. Initial states: $T_0 = \infty$ on the top plots and $ T_0 = 0 $ on the bottom plots. Vertical dashed line: $T_c = 2.269$. Label $\nu$ refers to the element $T_\nu$ of the flow Markov chain \eqref{eq:temp_flow}.}
\label{fig:RBM_flow_Alternative_Exp_Setups}
\end{figure*}

By feeding the machine with a MF training set, a flow towards $T_{c}= 2.269$ is unexpected. In this case the RBM has no information about the correct correlations of the system, neither about the right lattice geometry, since different lattice configurations can exhibit the same $z$. Nevertheless, even in this case, the NN thermometer produces a flow towards $T_{c}$. The results are shown in Figure \ref{fig:L010_RBM_MF_FLOW}.

The magnetization dynamics can also be analyzed in this scenario. Its fixed point is $m^{*} \approx 0.55$, considerably different from the one obtained in Figure \ref{fig:L010_RBM_ALL_mag}. However, due to its poor resolution near the fixed point, the thermometer still translates this result into a flow towards $T_c = 2.269$.

These results strongly suggest that we should be careful about using this particular experimental setup to check if RBMs capture relevant information from the dataset that can be connected to the RG. In this particular case the flow to the critical value $T_c = 2.269$ is artifactual. The machine has no information about the correct correlations of the model. Actually, only the measurement device does. 

\subsection{\label{sec:NO_MODEL} \texorpdfstring{$T = 0$}{T=0} and \texorpdfstring{$T = \infty$}{T=88} dataset }

 We now analyze another scenario where the RBM is fed with a dataset containing only perfectly ordered ($T=0$) and perfectly disordered ($T = \infty$) states. So that now the machine has no information about lattice topology or dimensionality. 

The resulting flows are shown in Figure \ref{fig:L100_RBM_ONLY_LOWHIGH}. The magnetization converges to  $m^{*} \approx 0.33$, what the thermometer again reads as  $T_c = 2.269$. Observe that, given the  NN thermometer precision, any map which introduces  order to initially random states ($T_0 = \infty$), or  disorder to ordered initial states will lead to temperature readings flowing towards $T_c$. 

Essentially, the only condition for the convergence to $T_c$ seems to be that $m^{*}$ lies in the range corresponding to the red region of the x-axis in Figure \ref{fig:L100_RBM_flow_mag_nearTc_MC}. The RBM identifies the order-disorder transition, while the correlations leading to $T_c = 2.269$ are actually captured by the thermometer.

Since the knowledge of the model is not necessary, perhaps this flow effect does not depend on the {\it quenched} weight values resulting from the training algorithm. In the next section we look into an {\it annealed} flow: at each Gibbs sampling iteration, the weights are sampled from a Gaussian distribution with appropriate mean and standard deviation.

\section{\label{sec:annealed} Annealed flow}

We have constructed histograms of the RBM coupling values after training throughout this work. Slightly shifted Gaussians were found in all simulations that led to flow towards $T_c$.

Following \cite{hinton_2012}, we initialized the RBM weight matrix sampling from a zero-mean Gaussian distribution with standard deviation $0.1$. Thus, if we only consider the distribution of couplings, the training procedure just shifts the initial Gaussian and rescales its standard deviation. That raises a question: do the detailed structure of couplings matter to produce the flow?

Remember from Figure \ref{FLOW_figure} that the RBM flow can be viewed as a multi-layer neural network with fixed weights. We then investigate the scenario where, instead of keeping fixed couplings, one samples them from a given distribution at each Markov chain step represented by \eqref{RBM_flow}. This setting is interesting because might allow the theoretical study of the flow by employing approaches introduced decades ago \cite{domany_1989}.

\begin{figure}[ht!]
\begin{center}
\centerline{\includegraphics[width=0.45\textwidth]{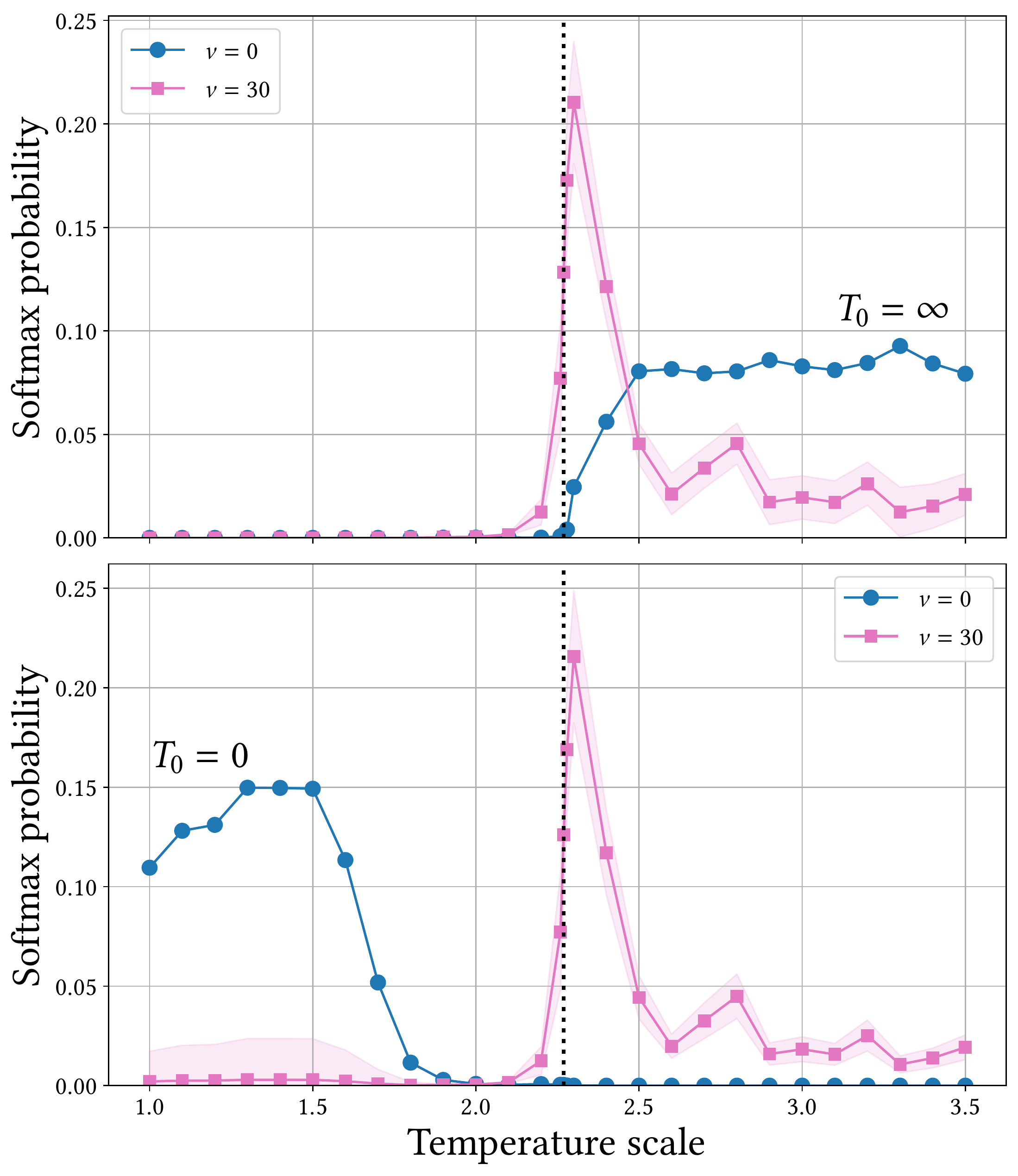}}
\caption{ Annealed version of the RBM flow ($L=100$). Solid lines: average over $50$ independent simulations. Filled colored areas: uncertainty within one standard deviation. Initial states: $T_0 = \infty$ on the top plot and $T_0= 0$ on the bottom plot. Vertical dashed line: $T_c= 2.269$. Label $\nu$ refers to the element $T_\nu$ of the flow Markov chain \eqref{eq:temp_flow}.}
\label{fig:L100_FLOW_ANNEALED}
\end{center}
\vskip -0.2in
\end{figure}

In Figure \ref{fig:L100_FLOW_ANNEALED} we present the fixed point averaged over $50$ independent simulations. The weights were sampled from a Gaussian distribution fitted over the histogram obtained in the scenario of Section \ref{sec:NO_MODEL}.

In addition to the possible theoretical path, this result also further strengthens the argument that the flow is artifactual. Again, the RBM flow tends to $T_c$, but the information about the model is fed only to the NN thermometer.

\section{\label{sec:w_matrix_analysis} Weight Matrix Analysis}

The next step to extract some meaningful information about what is happening is to  study the weight matrix $\bm{W}$. 

\subsection{\label{subsec:w_matrix_analysis_SVD} Singular Values of \texorpdfstring{$\bm{W}$}{W} }

We begin by calculating the singular value decomposition (SVD) of $\bm{W}$ after four different training situations:

\begin{enumerate}
  \item ${\cal T}_V$ dataset: temperatures from \eqref{eq:NN_scale_L010} for $L=10$ and from \eqref{eq:NN_scale_L100} for $L=100$. The flow tends to $T_c$.
  \item  $T=0$ dataset: only ordered states.  The fixed point is $T=0$. 
  \item  $T=\infty$ dataset: only random states. The fixed point is $T=\infty$.
  \item  $T=0$/$T=\infty$ dataset: only ordered and random states. The flow goes towards $T_c$.
\end{enumerate}

\begin{figure*}[htb!]
\centering
\begin{subfigure}[h]{.45\textwidth}
  \centering
  \includegraphics[width=\linewidth]{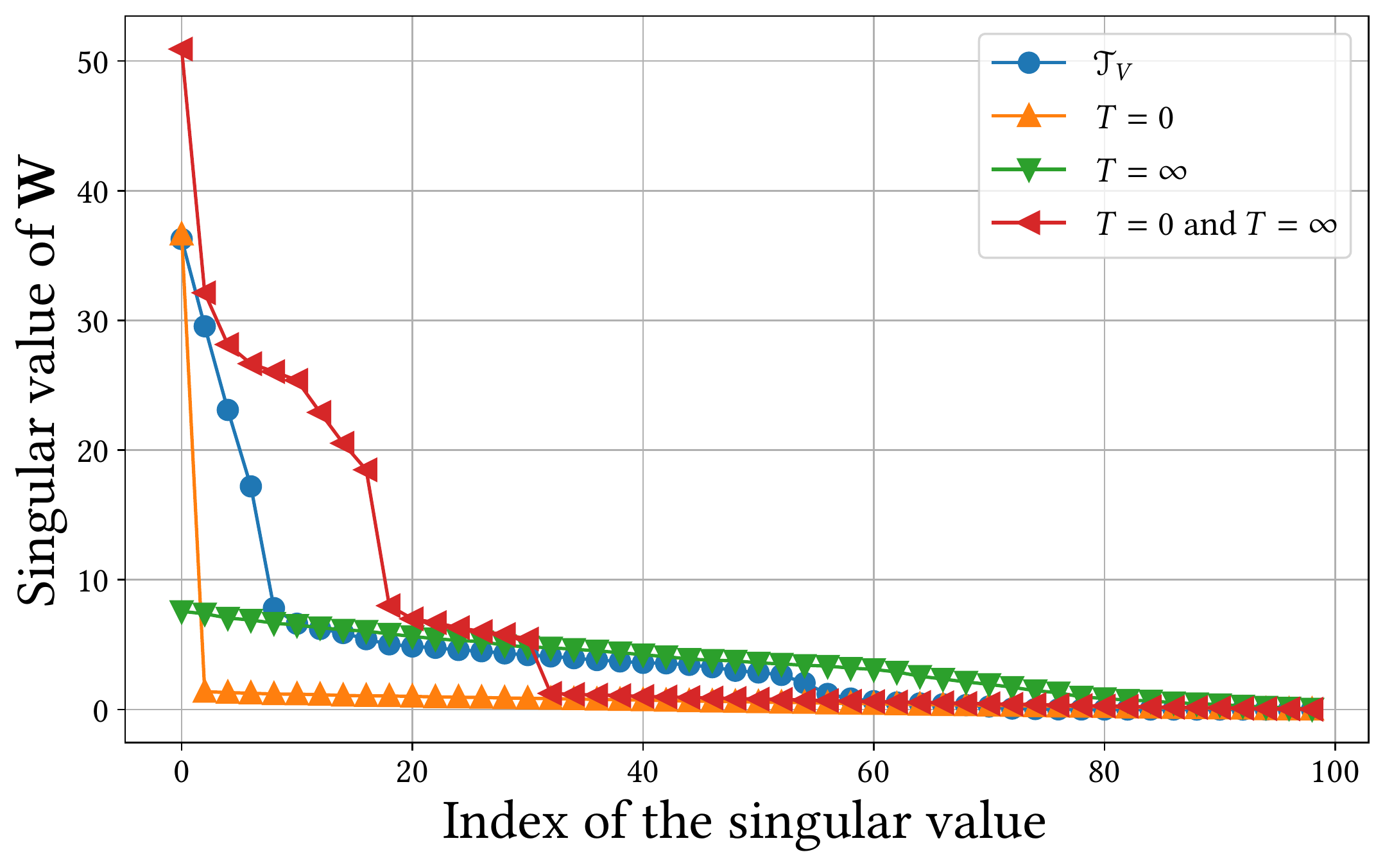}
  \caption{Singular values (descending order) of $\bm{W}$ for $L=10$.}
  \label{fig:L010_SVD}
\end{subfigure}%
\hspace{0.5cm}
\begin{subfigure}[h]{.45\textwidth}
  \centering
  \includegraphics[width=\linewidth]{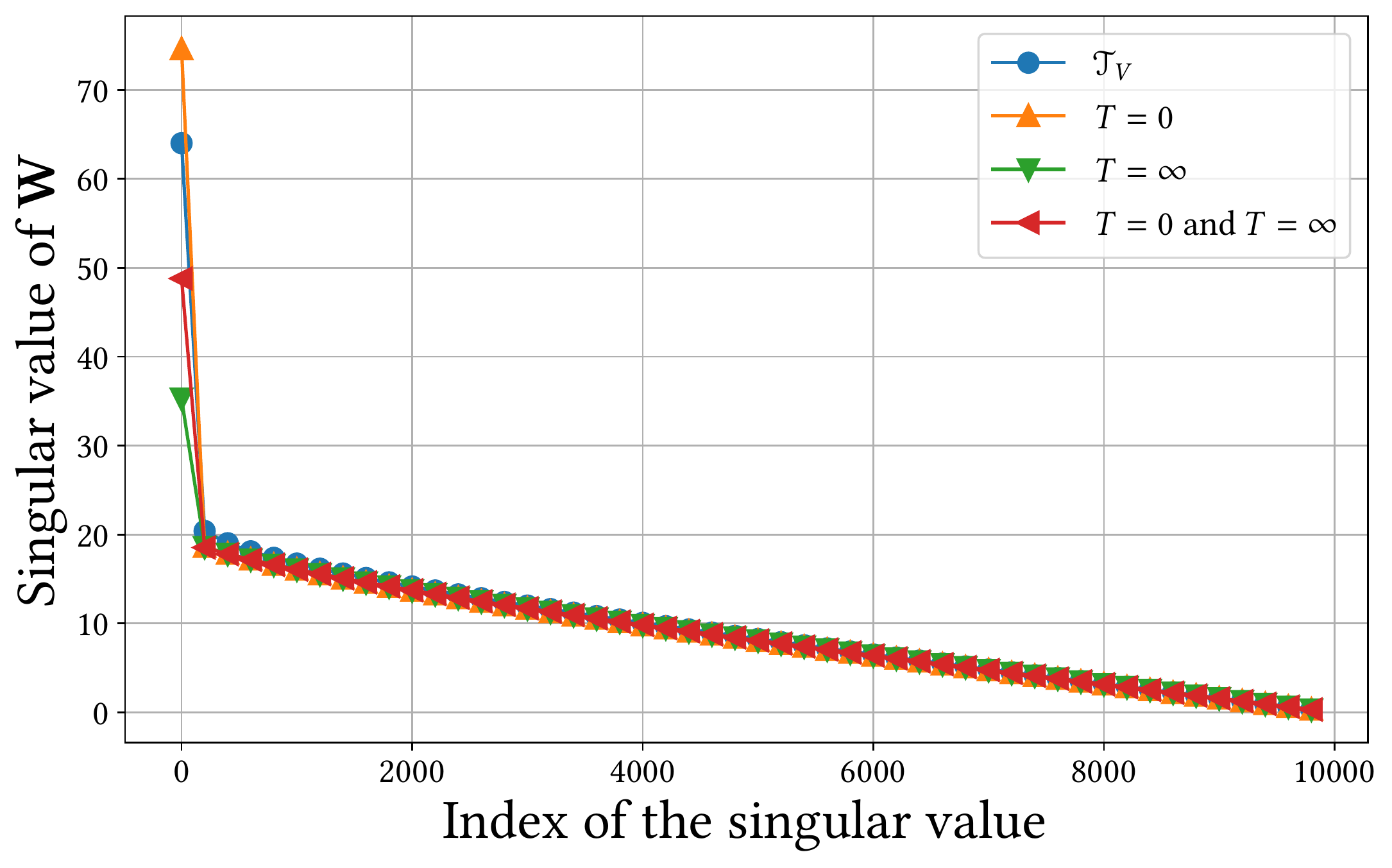}
  \caption{Singular values (descending order) of $\bm{W}$ for $L=100$}
 \label{fig:L100_SVD}
\end{subfigure}
\caption{Singular values of the weight matrix $\bm{W}$ for four different training scenarios.}
\end{figure*}

The singular values are presented in descending order for $L=10$ in Figure \ref{fig:L010_SVD}. There is a clearly distinguishable behavior for the cases where the flow goes towards $T_c$. Many singular values are relevant on those situations and `step' shape is observed. It is particularly interesting to note the complexity of the spectrum learned from a relatively simple dataset such as $T=0$/$T=\infty$. It indicates a highly nonlinear pattern and linear methods such as principal component analysis (PCA) could not be used to approximate the weight matrix.

The `step' shape suggests a way to verify whether the flow goes towards $T_c$ or not. However, once the lattice size is increased, the distinction is lost. The singular values are shown for $L=100$ in Figure \ref{fig:L100_SVD}.

\begin{figure}[ht!]
\begin{center}
\centerline{\includegraphics[width=0.45\textwidth]{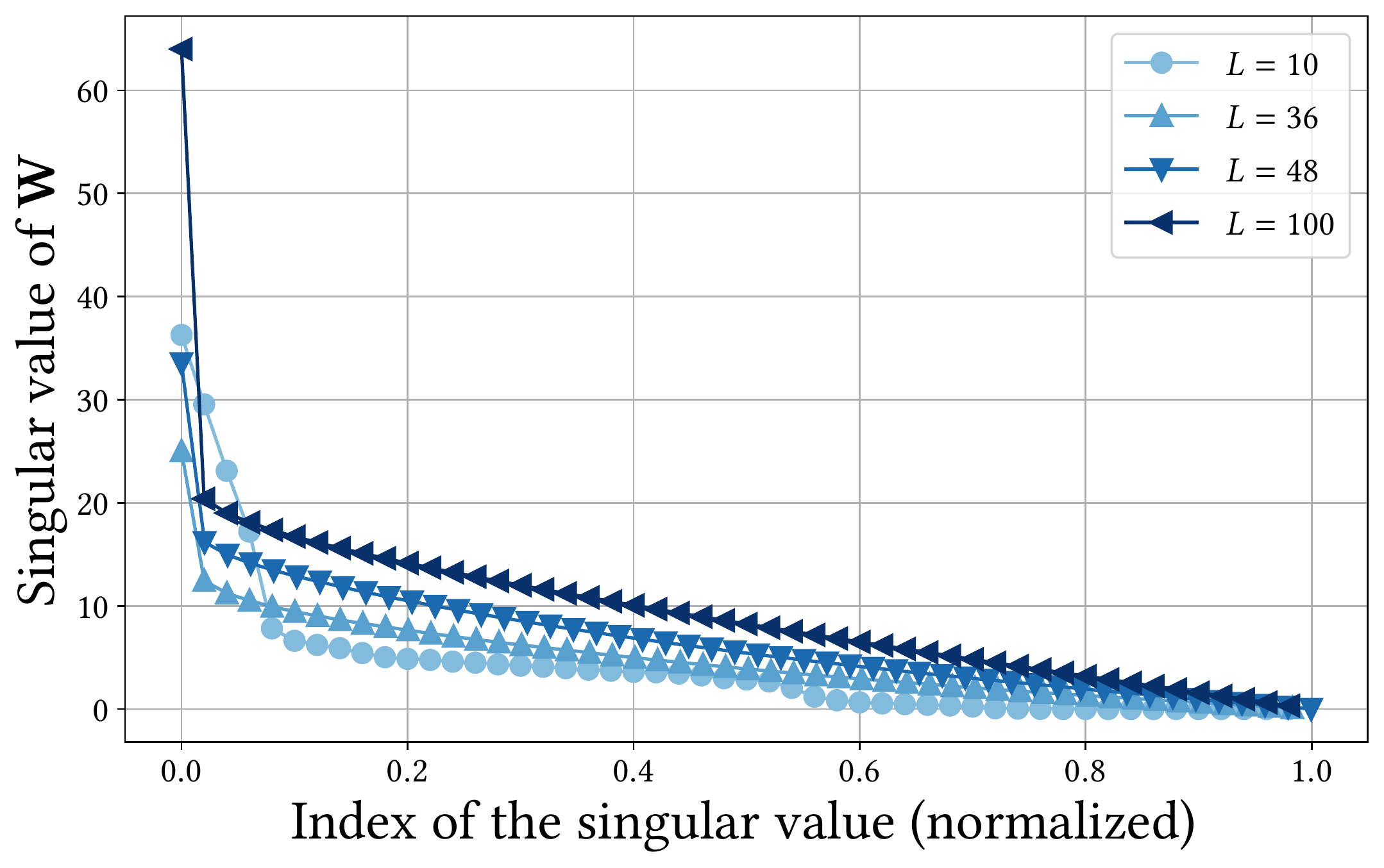}}
\caption{ Singular values (descending order) of $\bm{W}$ for different values of $L$. Dataset: ${\cal T}_V$.}
\label{fig:SVD_ALL}
\end{center}
\vskip -0.2in
\end{figure}

The singular spectrum is more complex for ${\cal T}_V$. Nonetheless, in contrast to $L=10$, the spectrum for $T=0$/$T=\infty$ looks more like $T=0$ and $T=\infty$ alone than ${\cal T}_V$, suggesting that no relevant information is learned by the RBM in this case. Since the flow still goes towards $T_c$, that finding bespeaks in favor to our warnings about the consequences of the resolution of the NN thermometer and the interpretation of the RBM flow in the temperature space. Additionally,  the `step' shape found for $L=10$ is actually a finite size effect. Indeed, we check that in Figure \ref{fig:SVD_ALL}.

\subsection{\label{subsec:w_matrix_analysis_SVS} Analysis of \texorpdfstring{$\bm{W} \bm{W}^{\top}$ }{WWt} }

Apart from the weight matrix itself, further studies can be done on the matrix $\bm{W} \bm{W}^{\top}$, which is a linear approximation for the correlations between units of the visible layer. Neglecting nonlinear contributions on Eqs.(\ref{eq:p_trans}), one finds an intra-layer interaction between visible units weighted by the elements of $\bm{W} \bm{W}^{\top}$.

\begin{figure*}[htb!]
\centering
\begin{subfigure}[h]{.45\textwidth}
  \centering
  \includegraphics[width=0.65\linewidth]{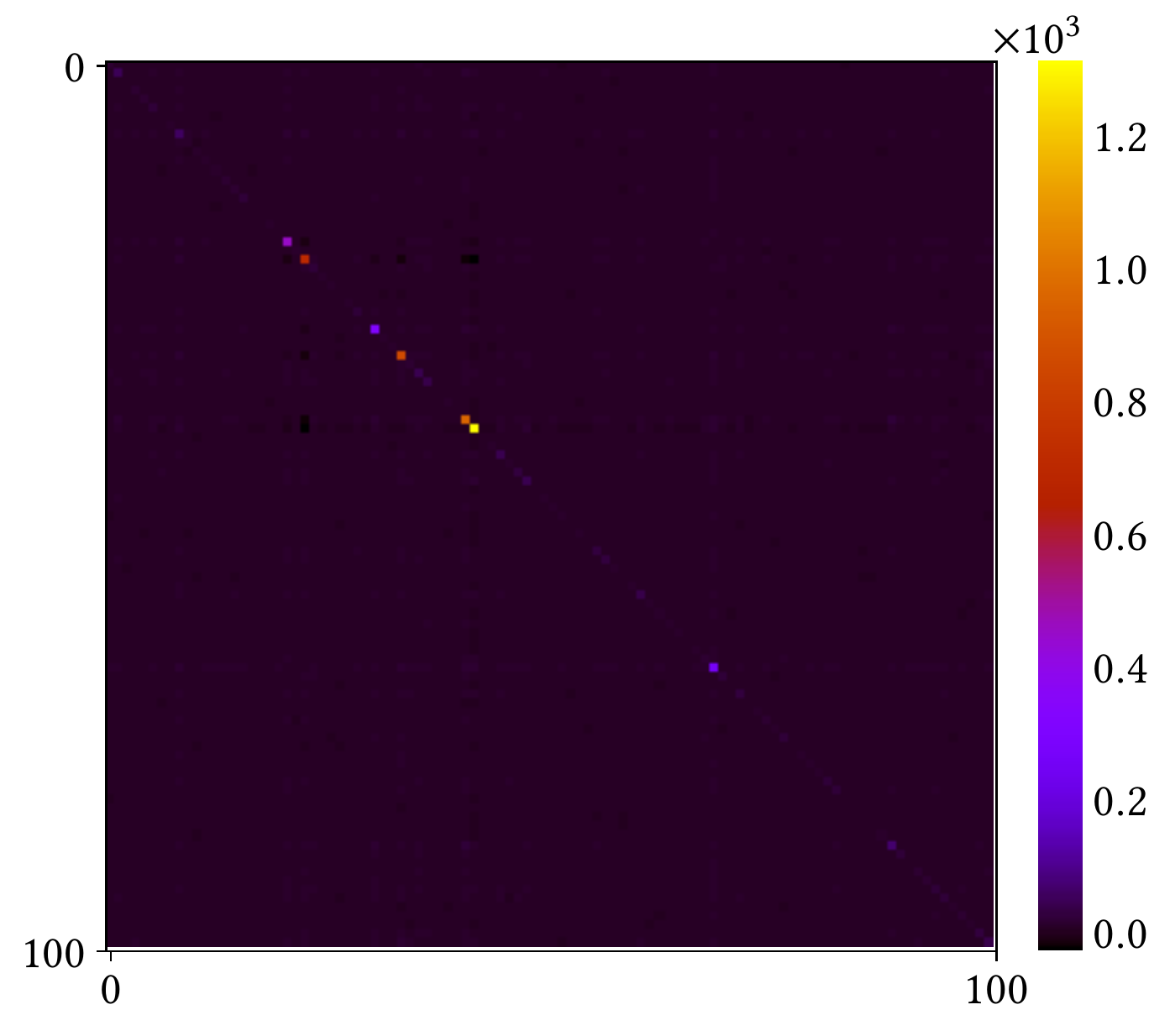}
  \caption{Dataset: ${\cal T}_V$.}
  \label{fig:L010_heatmapsA}
\end{subfigure}%
\hspace{0.5cm}
\begin{subfigure}[h]{.45\textwidth}
  \centering
  \includegraphics[width=0.65\linewidth]{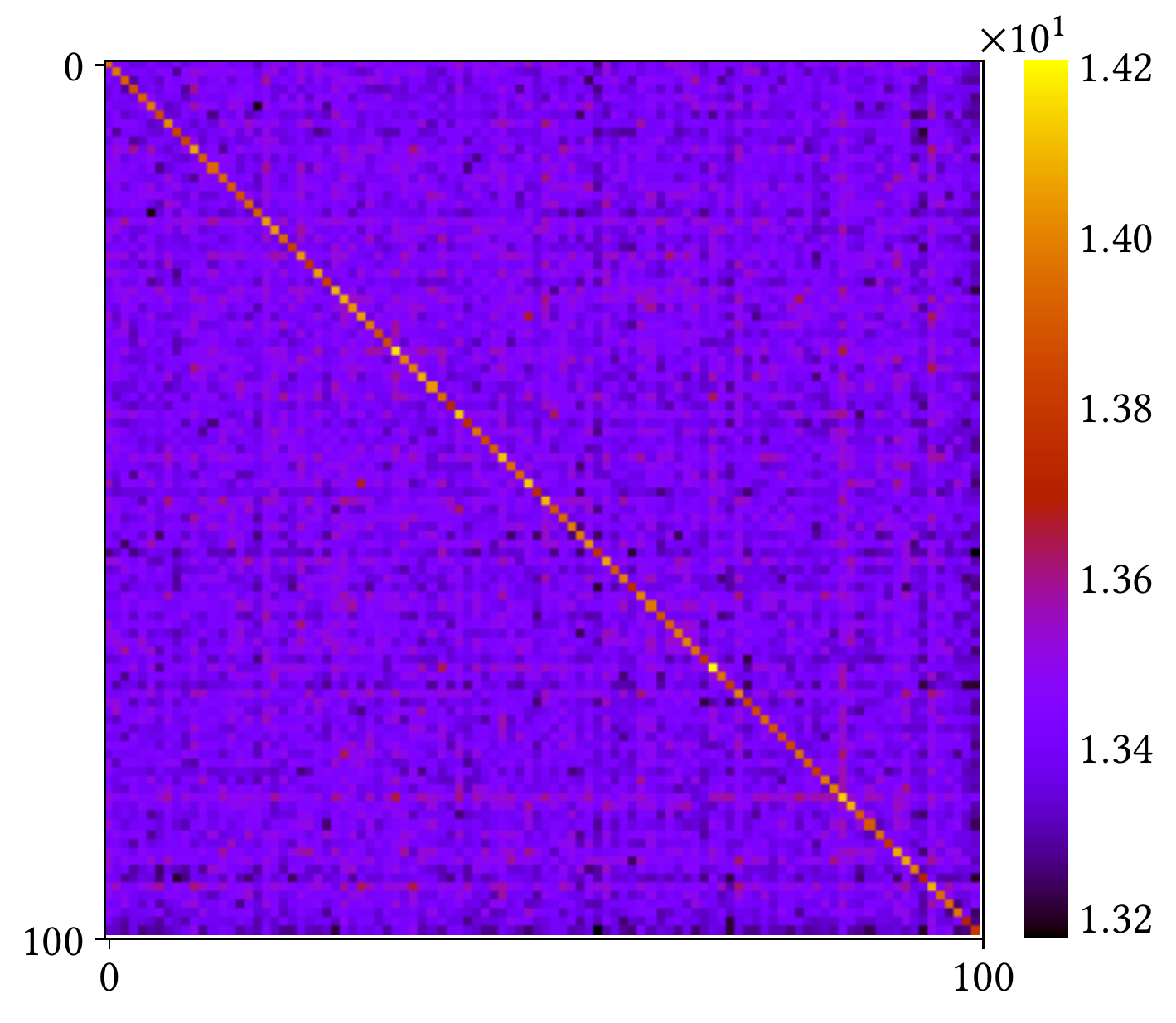}
  \caption{Dataset: $T=0$.}
 \label{fig:L010_heatmapsB}
\end{subfigure}
\begin{subfigure}[h]{.45\textwidth}
  \centering
  \includegraphics[width=0.65\linewidth]{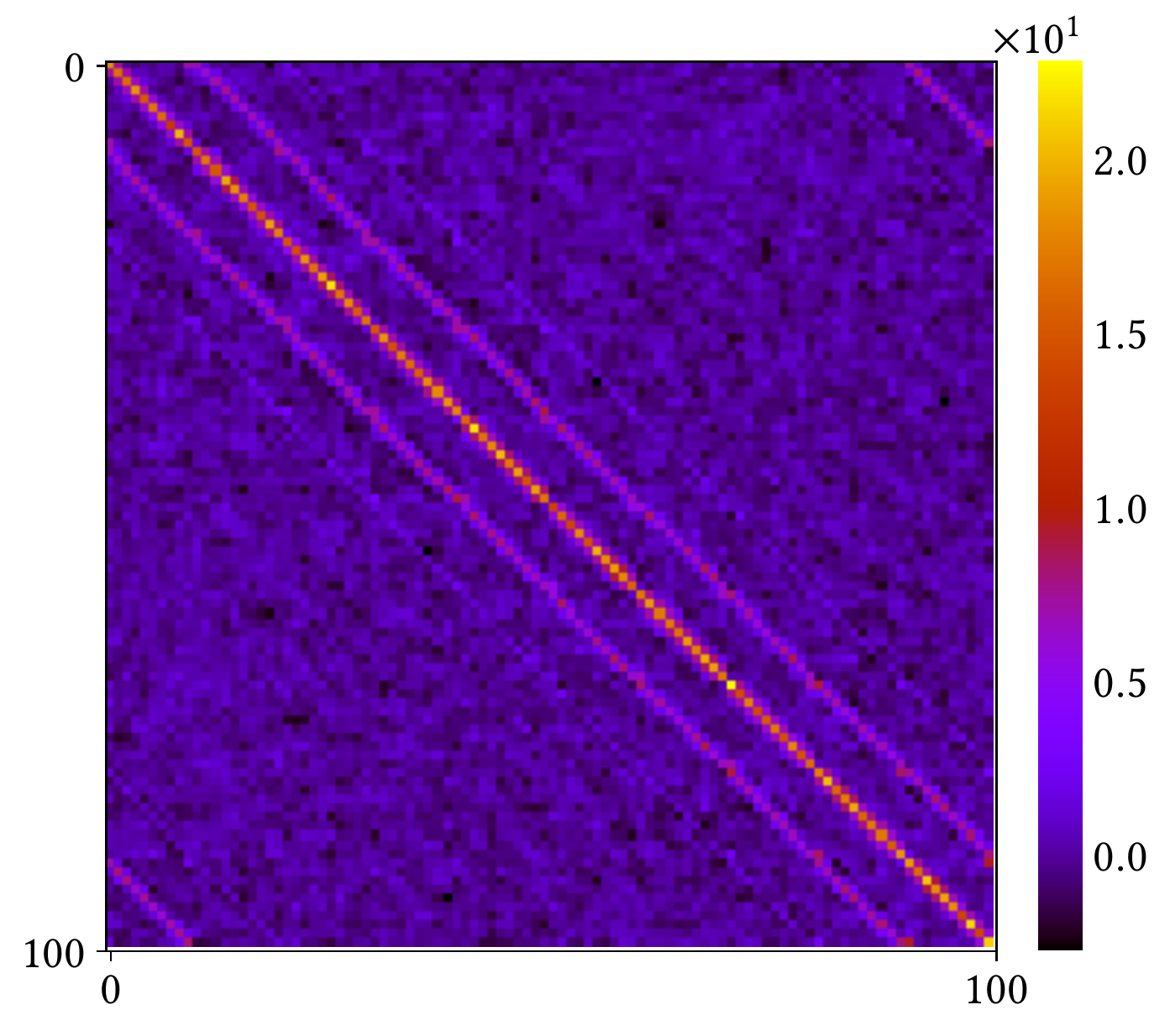}
  \caption{Dataset: $T=\infty$.}
 \label{fig:L010_heatmapsC}
\end{subfigure}%
\hspace{0.5cm}
\begin{subfigure}[h]{.45\textwidth}
  \centering
  \includegraphics[width=0.65\linewidth]{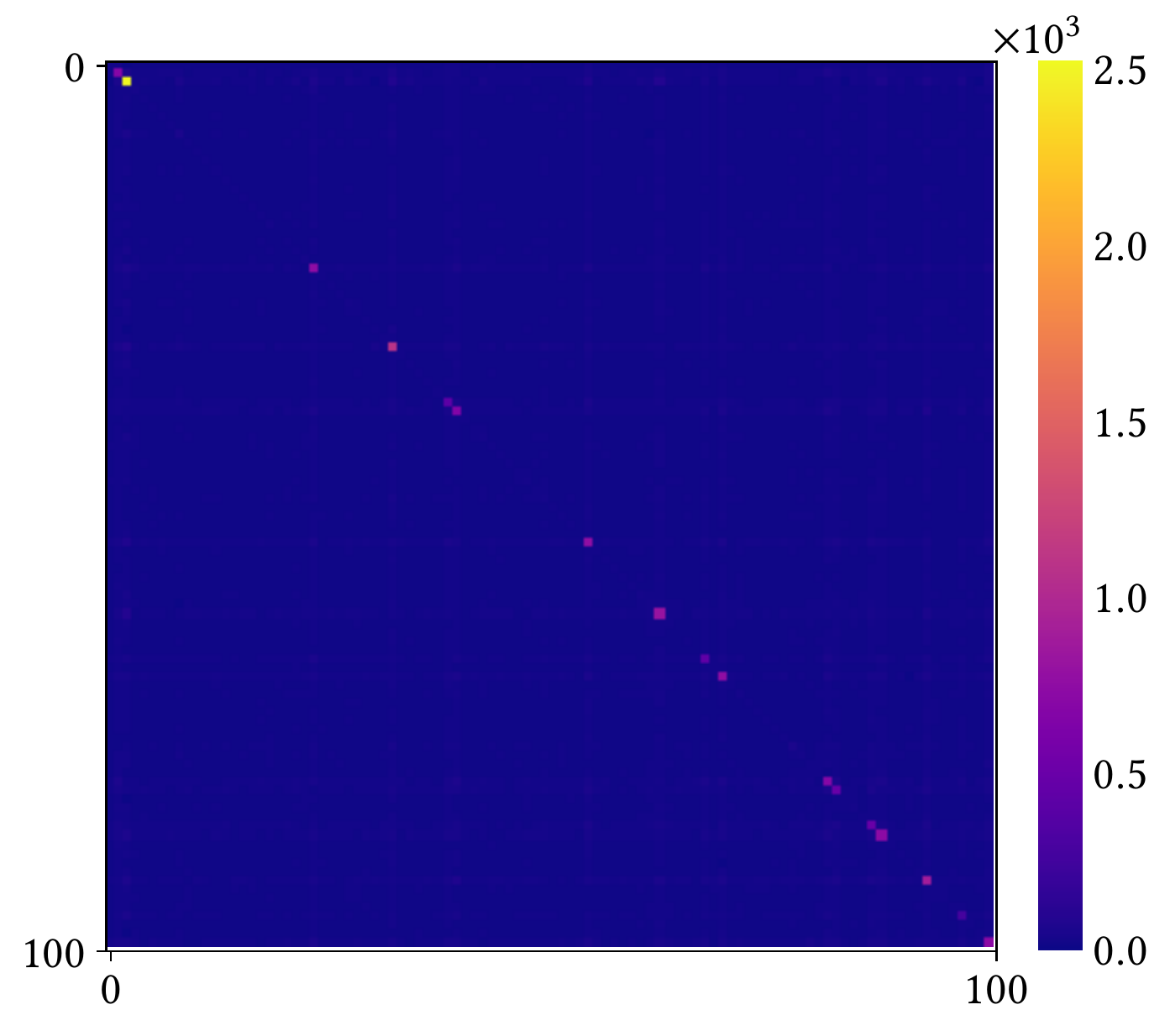}
  \caption{Dataset: $T=0$/$T=\infty$.}
 \label{fig:L010_heatmapsD}
\end{subfigure}
\caption{Visualizations of the matrix $\bm{W} \bm{W}^{\top}$ for $L=10$. }
\label{fig:L010_heatmaps}
\end{figure*}

In Figure \ref{fig:L010_heatmaps} we present for $L=10$ the matrix $\bm{W} \bm{W}^{\top}$ for the four training datasets. Similarly to the singular spectrum of Figure \ref{fig:L010_SVD}, the scenarios ${\cal T}_V$ and $T=0$/$T=\infty$ are similar. Both matrices are simple; just a few diagonal elements are non-zero ($\sim 10^3$). No interaction between neighboring sites ($(\bm{W} \bm{W}^{\top})_{jk}$ for $j\ne k$) is detected by the linear approximation for those cases, indicating that spin-spin correlations are mostly captured by non-linear terms. That suggests that  linearized RG transformations \cite{goldenfeld_1992} are not enough to study the RBM flow problem.

On the other hand, though the presence of dominant diagonal elements ($\sim 10$), matrices for $T=0$ and $T=\infty$ have multiple interacting sites contributions. These contributions, at least within linear approximation, can be regarded as `noise' to the RBM flow, since it does not go towards $T_c$. 

\begin{figure*}[htb!]
\centering
\begin{subfigure}[h]{.225\textwidth}
  \centering
  \includegraphics[width=0.9\linewidth]{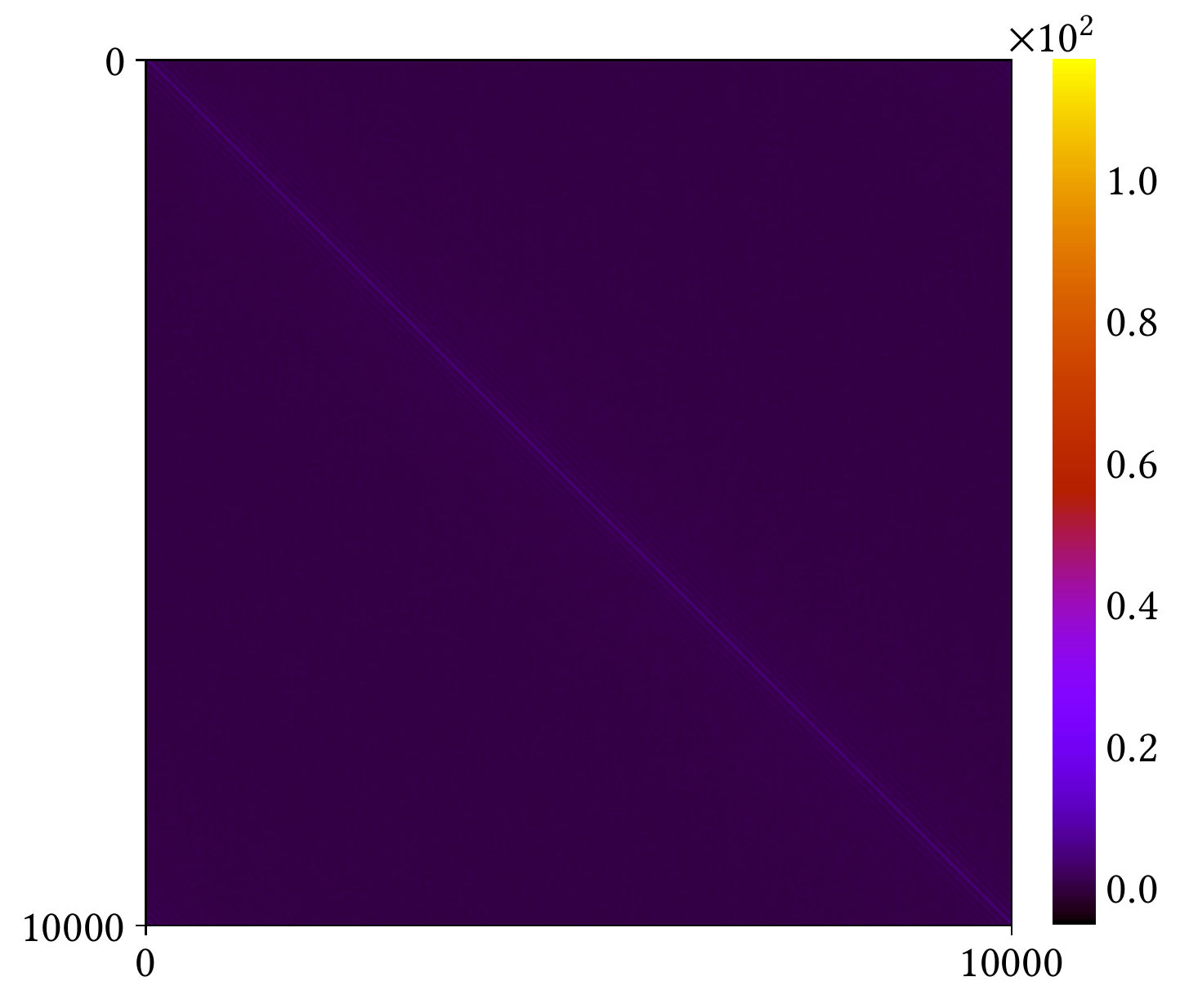}
  \caption{Dataset: ${\cal T}_V$.}
  \label{fig:L100_heatmapsA}
\end{subfigure}%
\hspace{0.25cm}
\begin{subfigure}[h]{.225\textwidth}
  \centering
  \includegraphics[width=0.9\linewidth]{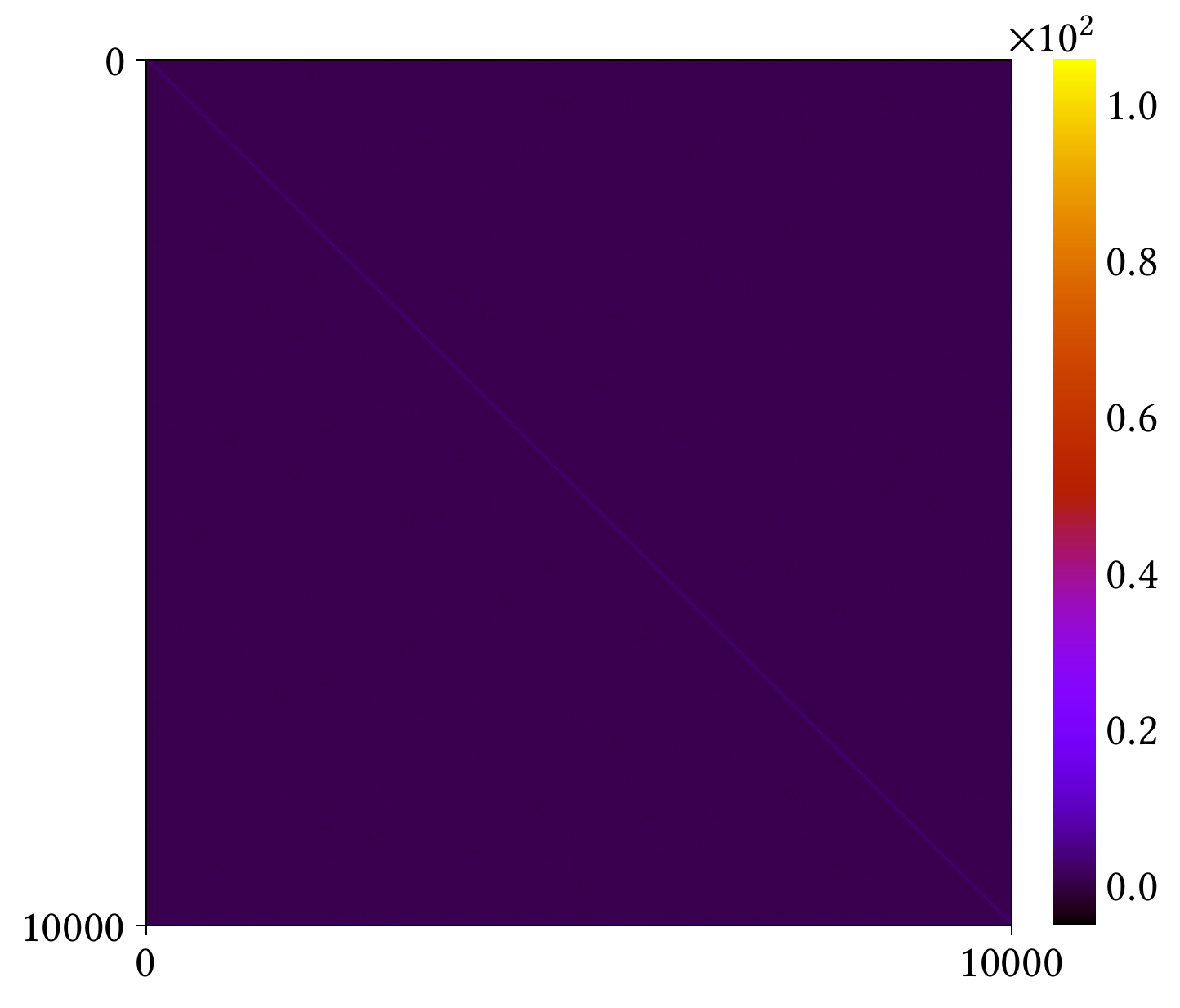}
  \caption{Dataset: $T=0$.}
 \label{fig:L100_heatmapsB}
\end{subfigure}%
\hspace{0.2cm}
\begin{subfigure}[h]{.225\textwidth}
  \centering
  \includegraphics[width=0.9\linewidth]{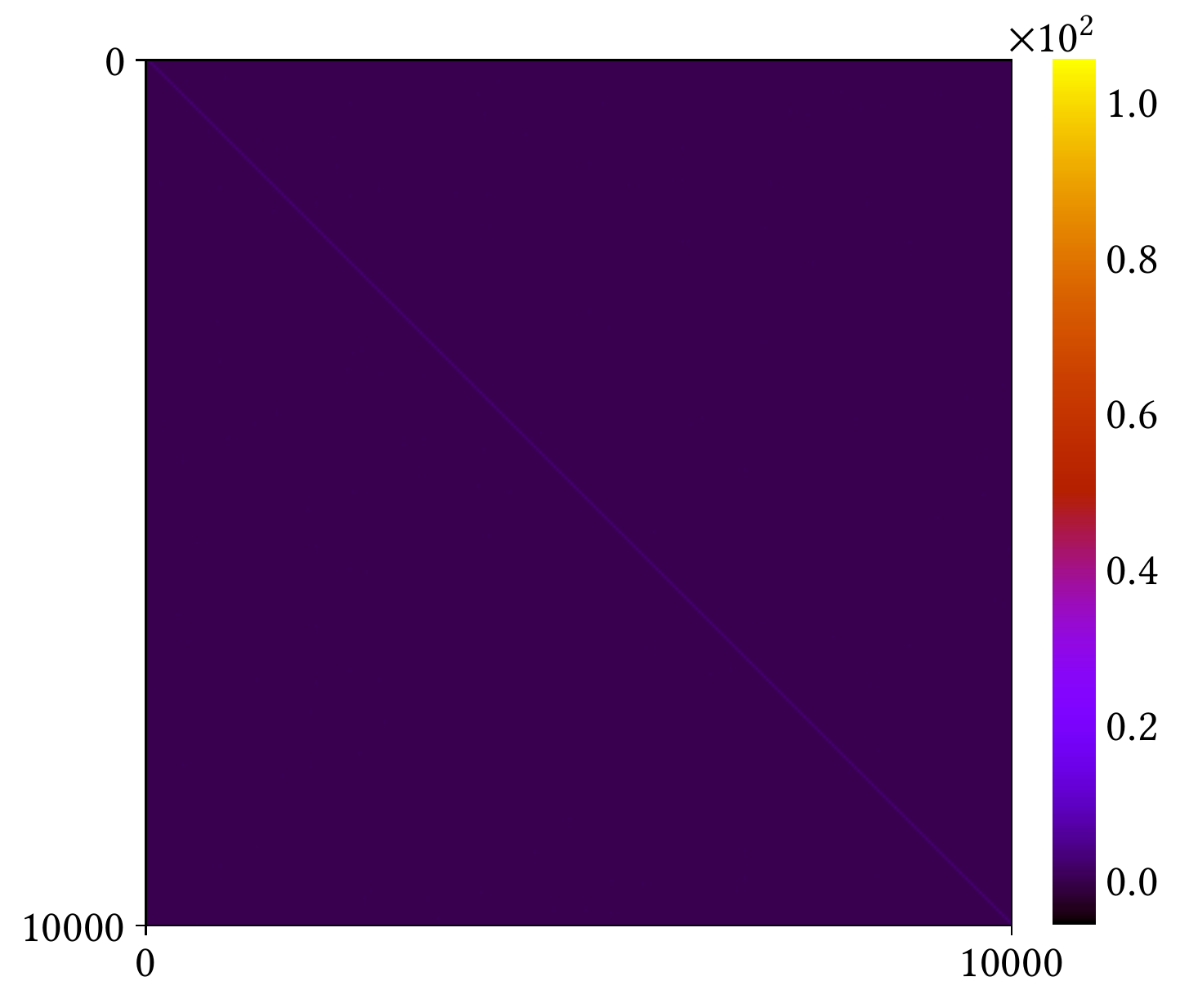}
  \caption{Dataset: $T=\infty$.}
 \label{fig:L100_heatmapsC}
\end{subfigure}%
\hspace{0.2cm}
\begin{subfigure}[h]{.225\textwidth}
  \centering
  \includegraphics[width=0.9\linewidth]{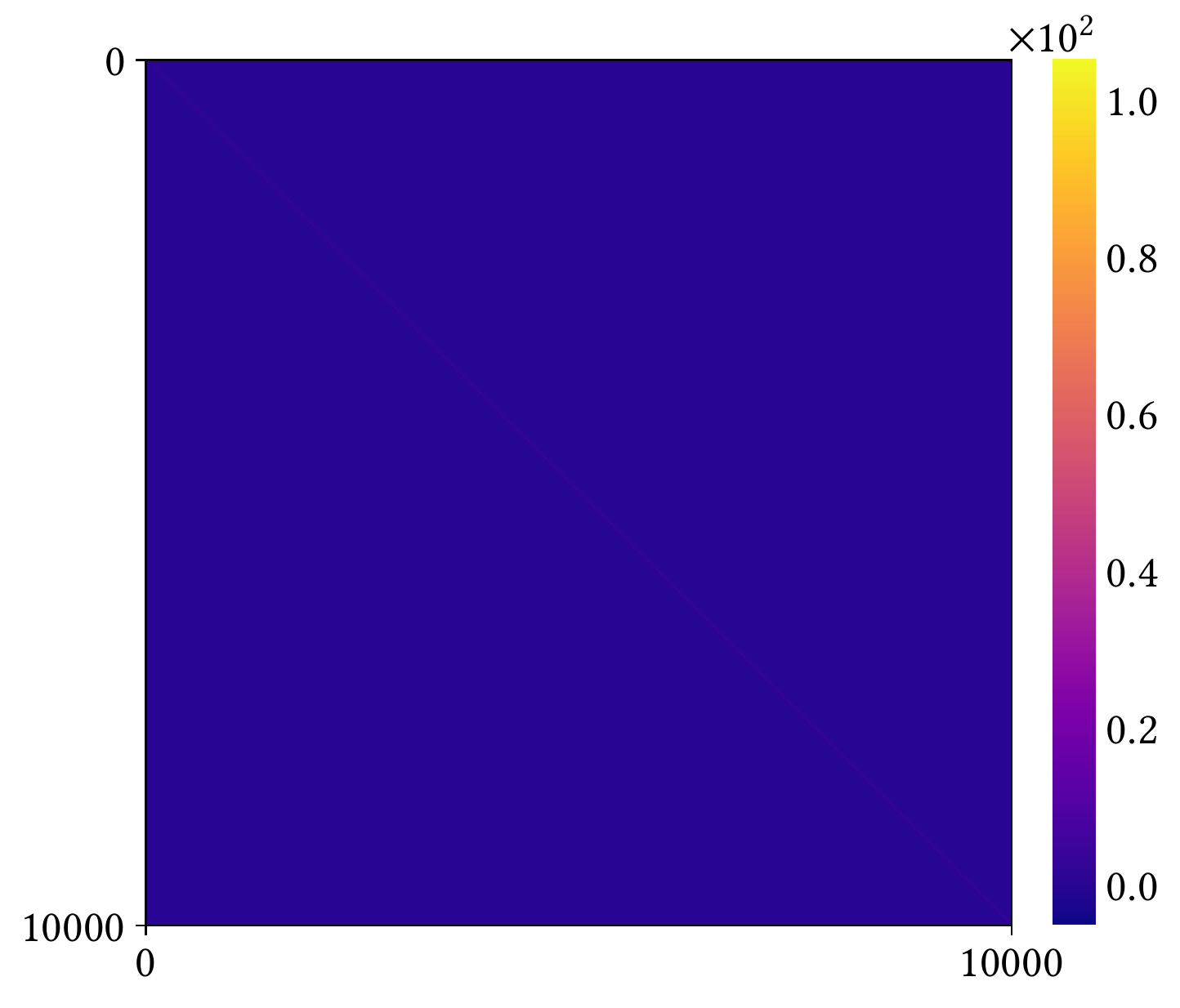}
  \caption{Dataset: $T=0 / T = \infty$.}
 \label{fig:L100_heatmapsD}
\end{subfigure}
\begin{subfigure}[h]{.225\textwidth}
  \centering
  \includegraphics[width=0.9\linewidth]{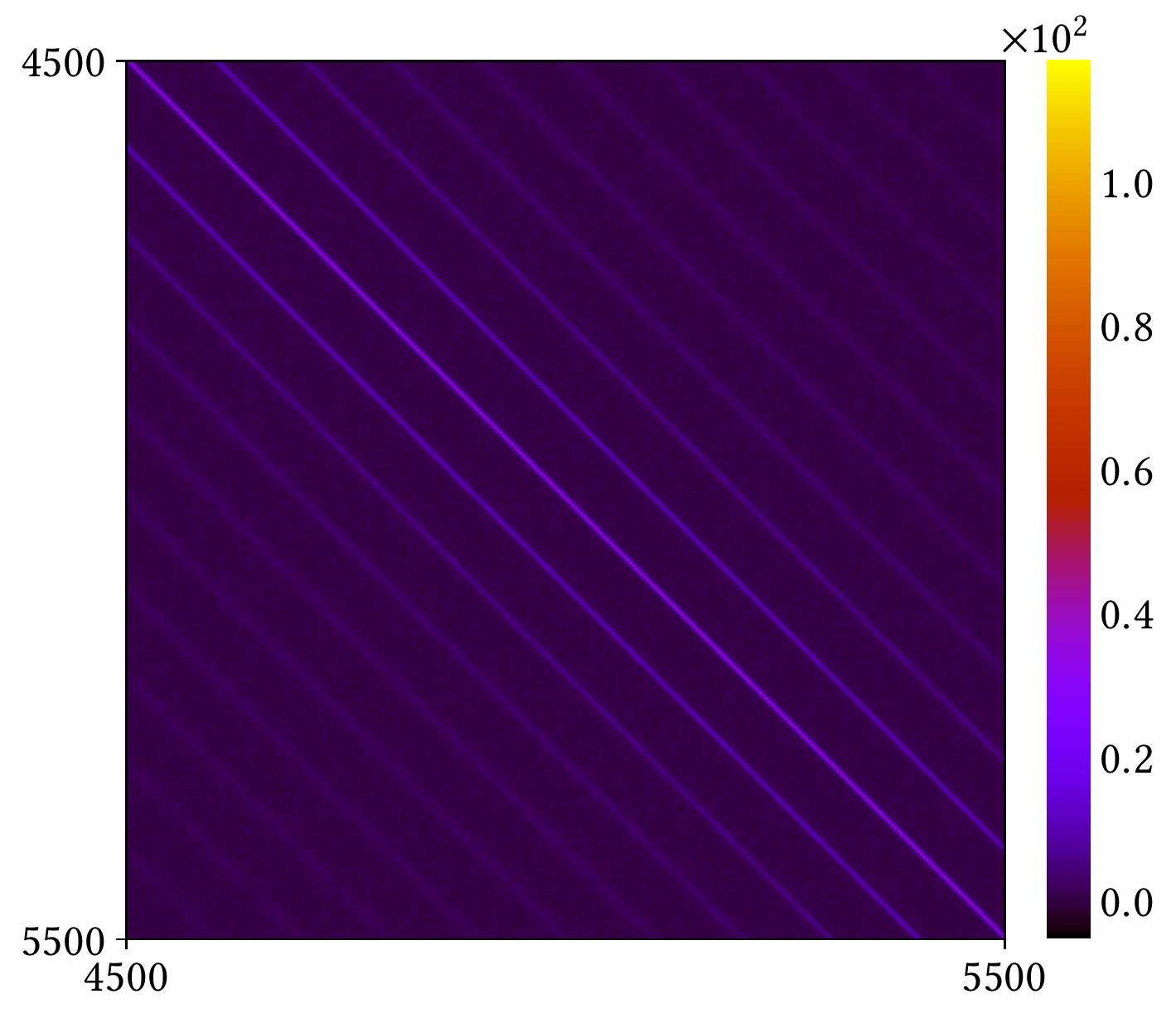}
  \caption{Zoom \\ Dataset: ${\cal T}_V$.}
  \label{fig:L100_heatmaps_zoomA}
\end{subfigure}%
\hspace{0.2cm}
\begin{subfigure}[h]{.225\textwidth}
  \centering
  \includegraphics[width=0.9\linewidth]{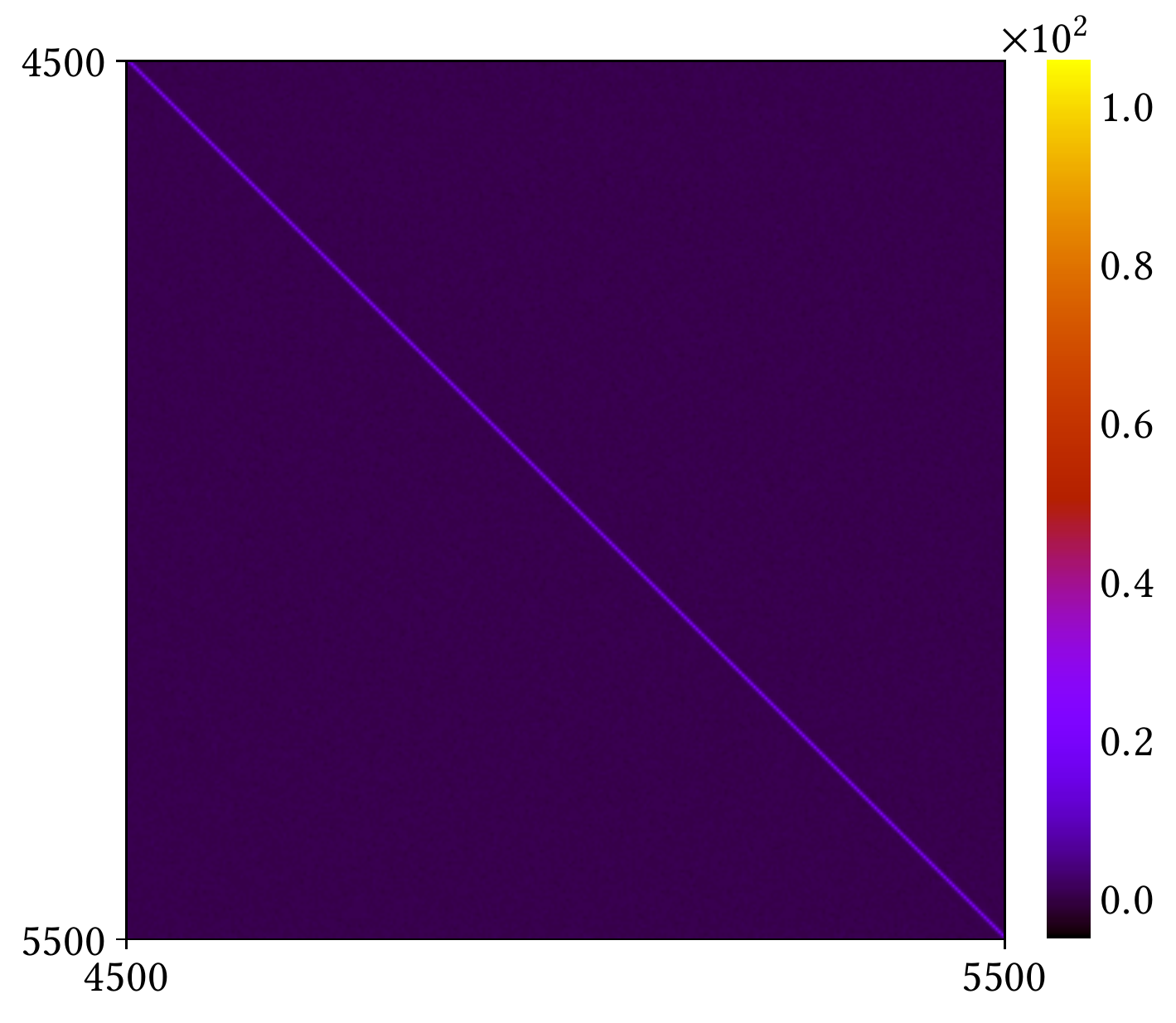}
  \caption{Zoom \\ Dataset: $T=0$.}
 \label{fig:L100_heatmaps_zoomB}
\end{subfigure}%
\hspace{0.2cm}
\begin{subfigure}[h]{.225\textwidth}
  \centering
  \includegraphics[width=0.9\linewidth]{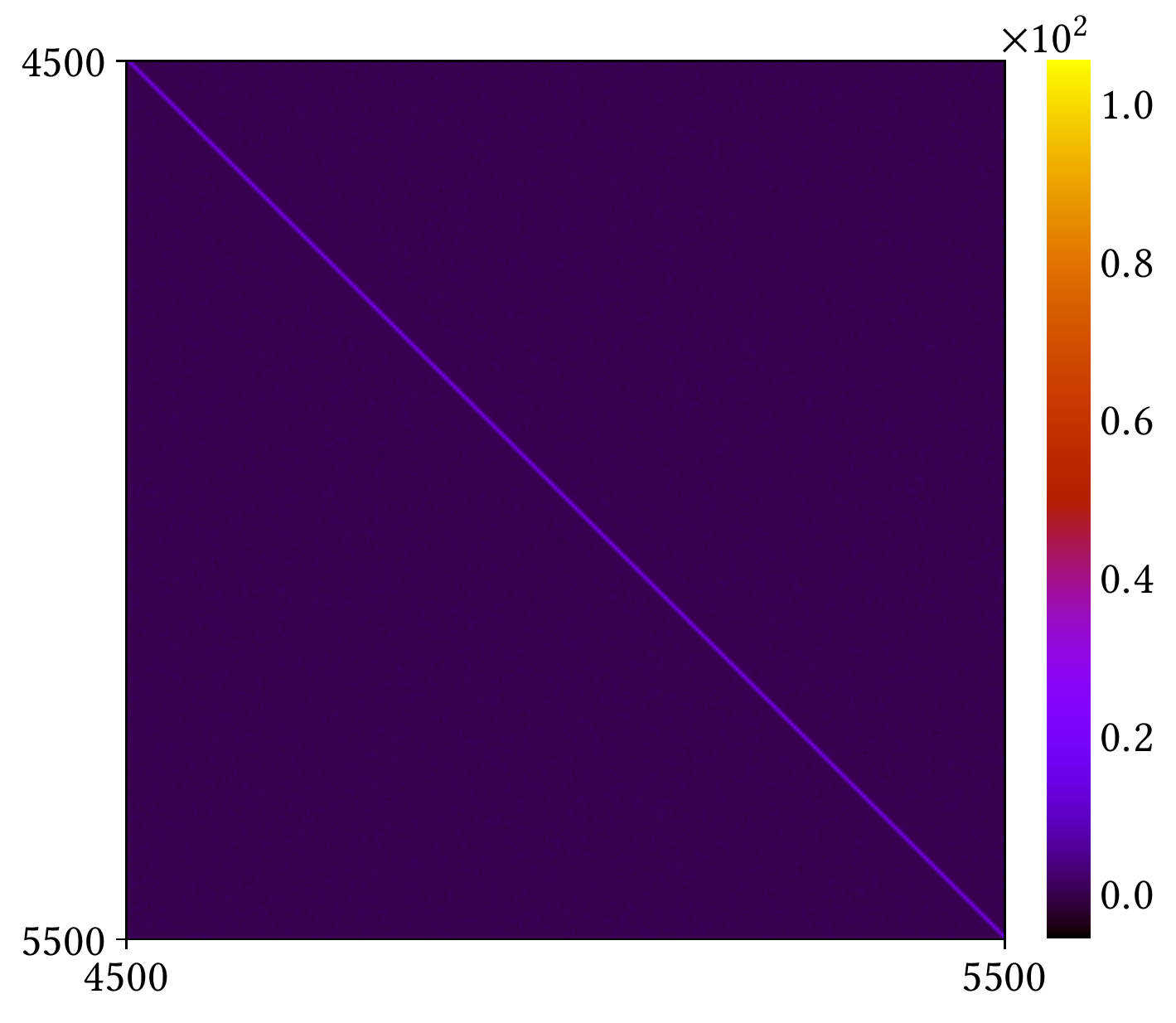}
  \caption{Zoom \\ Dataset: $T= \infty$.}
 \label{fig:L100_heatmaps_zoomC}
\end{subfigure}%
\hspace{0.2cm}
\begin{subfigure}[h]{.225\textwidth}
  \centering
  \includegraphics[width=0.9\linewidth]{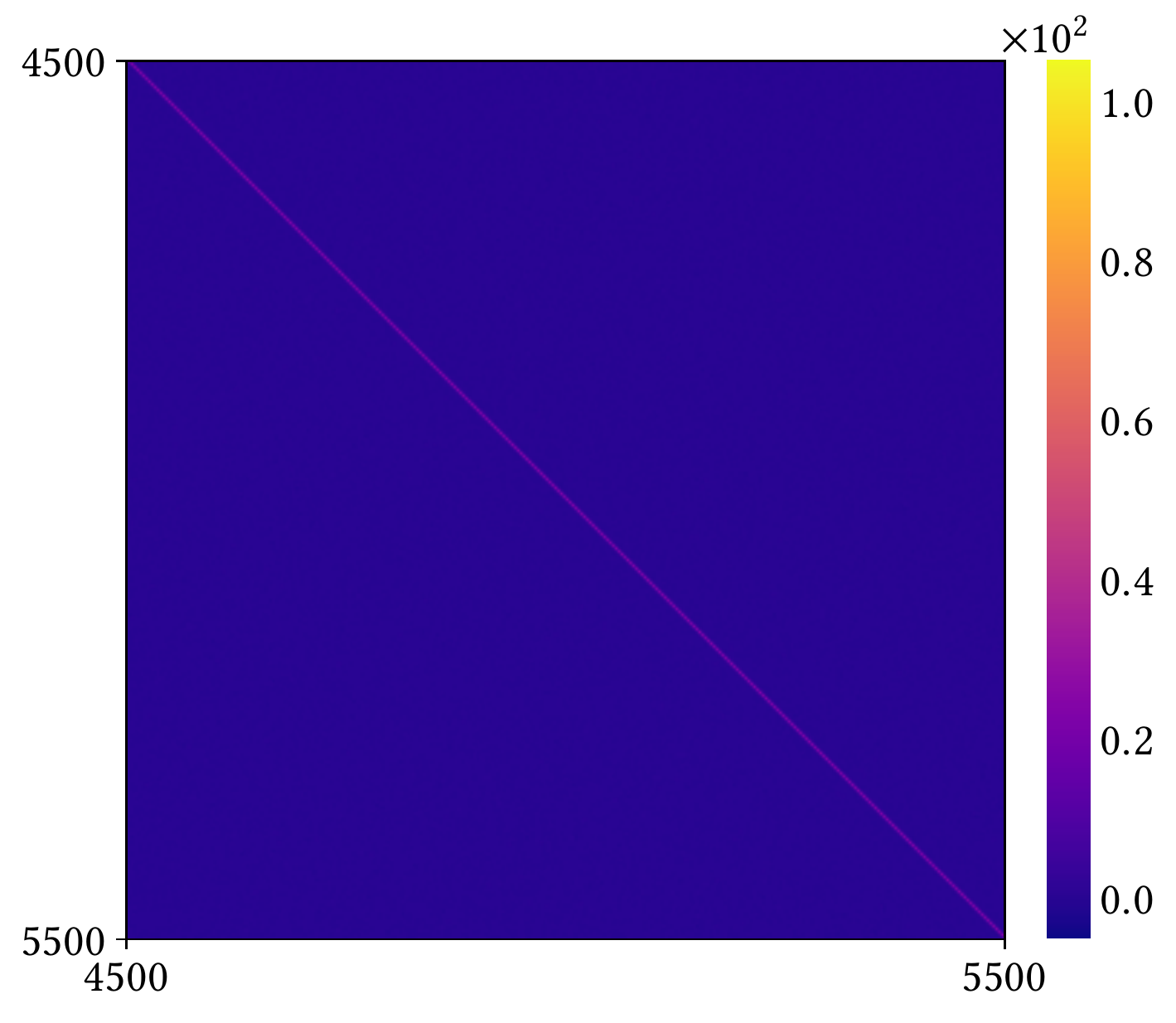}
  \caption{Zoom \\ Dataset: $T=0 / T = \infty$.}
 \label{fig:L100_heatmaps_zoomD}
\end{subfigure}
\caption{Visualizations of the matrix $\bm{W} \bm{W}^{\top}$ for $L=100$. Figures \ref{fig:L100_heatmaps_zoomA}-\ref{fig:L100_heatmaps_zoomD} are zooms of Figures \ref{fig:L100_heatmapsA}-\ref{fig:L100_heatmapsD}, respectively. }
\label{fig:L100_heatmaps_fig}
\end{figure*}

Similarly to the `step' behavior of the singular spectrum, correlation patterns shown in $\bm{W} \bm{W}^{\top}$ could indicate whether the flow goes tends $T_c$ or not. However there is no clear distinction between the four scenarios when the system size is $L=100$ as it is shown in Figures \ref{fig:L100_heatmapsA}-\ref{fig:L100_heatmapsD}.

In Figures \ref{fig:L100_heatmaps_zoomA}-\ref{fig:L100_heatmaps_zoomD} we take a closer look. The RBM captures some correlation between neighbouring sites (non-diagonal elements) for ${\cal T}_V$. However, the training with $T=0$/$T=\infty$ has not detected these correlations in the linear approximation but still produced a flow towards $T_c$.

\begin{figure*}[htb!]
\centering
\begin{subfigure}[h]{.45\textwidth}
  \centering
  \includegraphics[width=\linewidth]{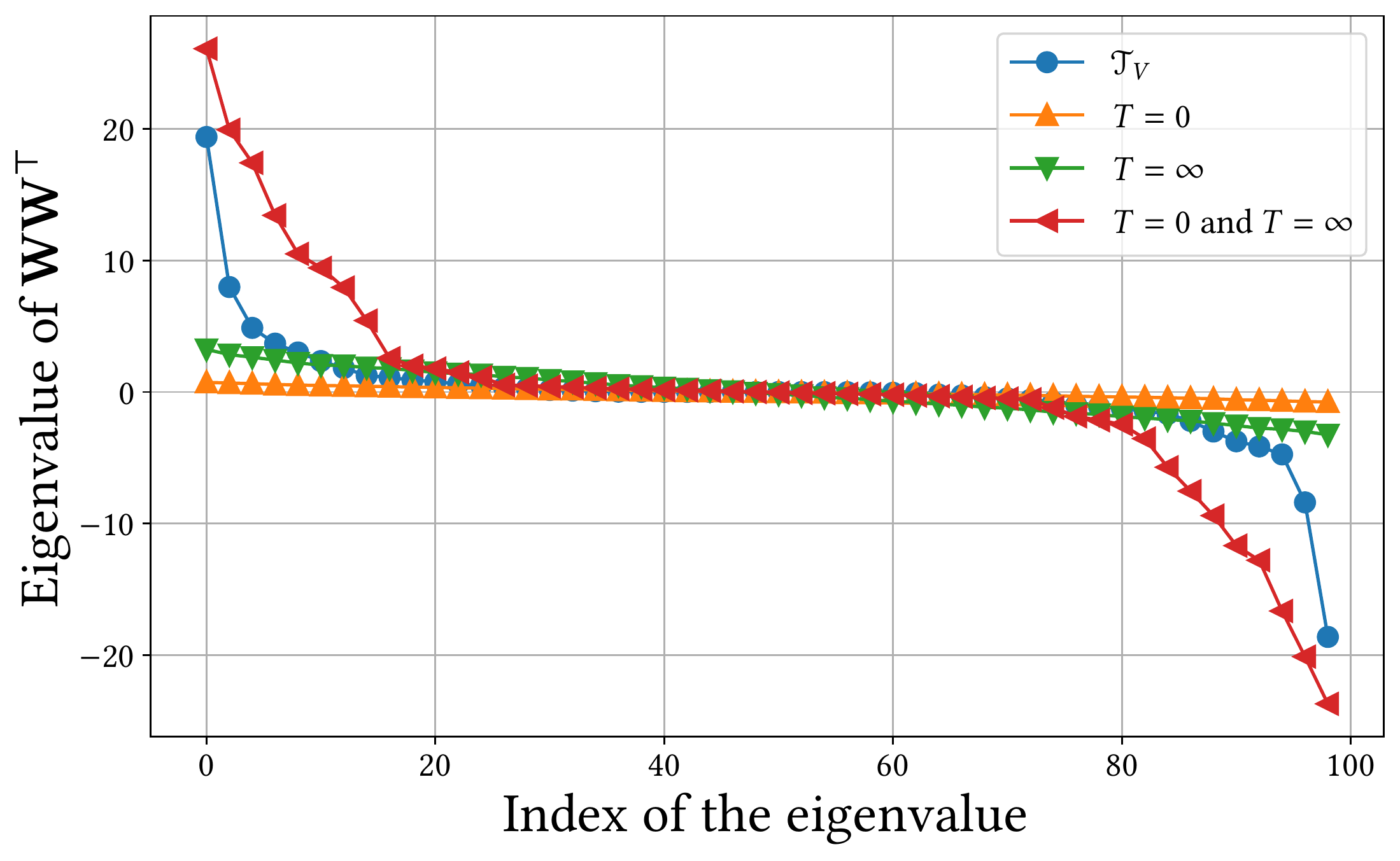}
  \caption{Eigenvalues (descending order) of $\bm{W} \bm{W}^{\top}$ for $L=10$.}
  \label{fig:L010_Eigen}
\end{subfigure}%
\hspace{0.5cm}
\begin{subfigure}[h]{.45\textwidth}
  \centering
  \includegraphics[width=\linewidth]{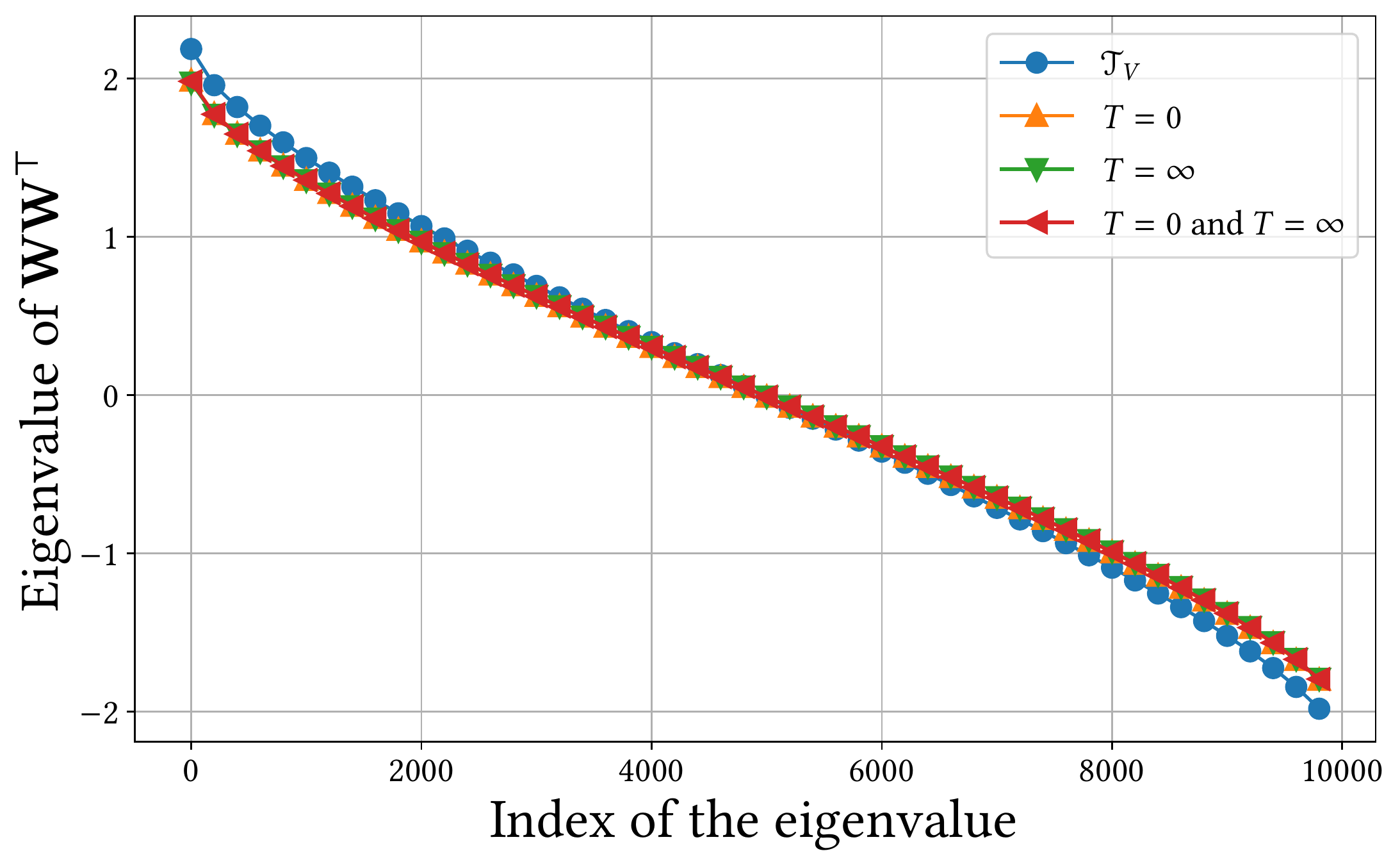}
  \caption{Eigenvalues (descending order) of $\bm{W} \bm{W}^{\top}$ for $L=100$.}
 \label{fig:L100_Eigen}
\end{subfigure}
\caption{Eigenvalues of the matrix $\bm{W} \bm{W}^{\top}$ for four different training situations.}
\end{figure*}

To conclude, we take a look at the eigenvalues of $\bm{W} \bm{W}^{\top}$. Similarly to the `step' shape on the singular spectrum values characterizing the flow for $L=10$, there is a distinctive `gap' on the eigenvalue spectrum which seems to close when flows do not tend to $T_c$. This observation is shown in Figure \ref{fig:L010_Eigen}.

\begin{figure}[ht!]
\begin{center}
\centerline{\includegraphics[width=0.45\textwidth]{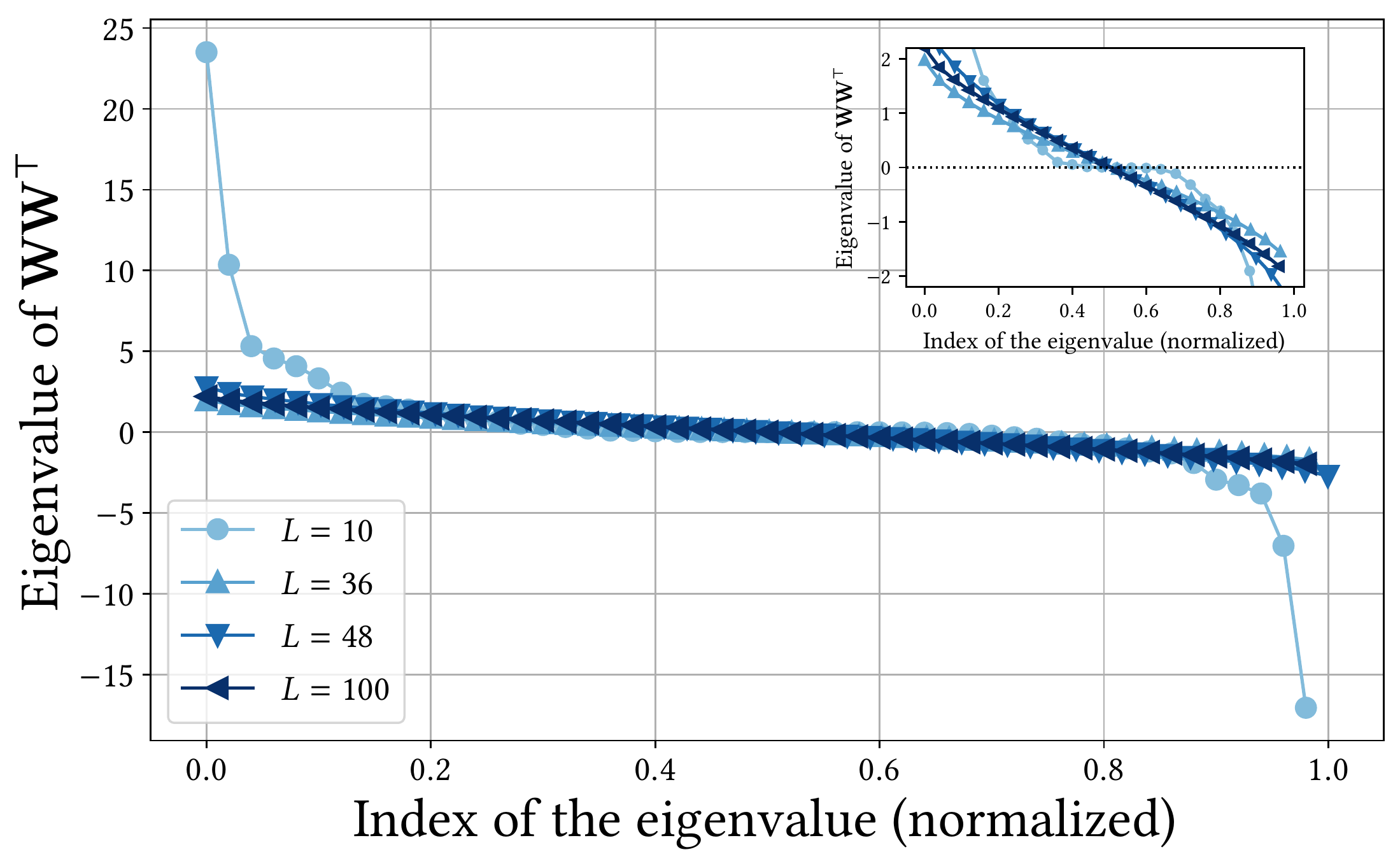}}
\caption{Eigenvalues (descending order) of $\bm{W} \bm{W}^{\top}$ for different values of $L$. Dataset: ${\cal T}_V$. Inset: zoom over eigenvalues near zero.}
\label{fig:Eigen_ALL}
\end{center}
\vskip -0.2in
\end{figure}

However, for $L=100$ the spectrum for all datasets looks similar as shown in Figure \ref{fig:L100_Eigen}. Again, the features obtained for $L=10$ seem to be a finite-size effect as it can be seen in Figure \ref{fig:Eigen_ALL}.

\section{\label{sec:conclusion} Concluding Remarks}

A connection between feature extraction in an standard unsupervised learning setting and RG would be an important achievement. There would be benefits both for the theoretical understanding of ML models (still lacking in many cases, such as Deep Learning) and for ML practitioners and physicists who have incorporated ML algorithms in their toolboxes. The well established RG framework could give several hints to improve performance and speed-up training.

Intuitively, it is quite natural to relate some ML models with the iterative hierarchical coarse-graining procedure of RG, that extracts relevant information from systems involving many scales \cite{shwartz-ziv_2017}. In order to investigate connections between RBMs and RG proposed by Mehta and Schwab \cite{mehta_2014}, Iso {\it et al.} \cite{iso_2018} introduced an NN thermometer and  used it to conclude that the RBM flow could be used to find the critical temperature. 

Here, in order to test whether the machine is really learning scale invariant features from a multi-temperature dataset, we have proposed alternative numerical experiments. First, we have analyzed the RBM flow resulting from training the machine in samples from a mean field model, while employing the original NN thermometer, to find the same $T_c=2.269$. We then trained the RBM in a dataset composed by totally ordered and totally disordered samples to, yet again, use the same NN thermometer to  find  $T_c = 2.269$. We then proceed by using a RBM with random weights with the same distribution of the original trained machine. Again finding the RBM flow and $T_c$. This series of numerical experiments suggest that the relevant geometrical information was learned by the NN thermometer, while the RBM captured the existence of two phases. Finally, information about the transition is encoded in the matrix of weights. However, this information is encapsulated on  non-linear terms of the machine and is hard to detect when the lattice size increases, suggesting that linearized RG transformations are not enough to study the flow problem.

In conclusion, the experiments we have performed show that one has to be careful using a NN thermometer as part of a setup designed to discuss the detection of scale invariance. The analysis of the possible connection between standard RBMs and scale invariance, or RG, remains elusive. We think it could benefit from other thermometer designs, perhaps in a Bayesian scenario, and from the study of the flow as a probabilistic dynamical system. 

\paragraph{Update remark} Recently, a study of the RBM flow without using the
NN thermometer has been proposed in Ref. \cite{funai_2021}.

\section*{Acknowledgements}
This work had financial support from the Brazilian National Council for Scientific and Technological Development (CNPq) under process 162857/2017-9. We thank Nestor Caticha for fruitful discussions and for pointing our attention to Ref. \cite{domany_1989}. We also give thanks to Carlos Neves for inspiring discussions.
\newpage

\bibliographystyle{IEEEtran}
\bibliography{RBMflow}

\end{document}